\documentclass[aps,prb,floatfix,twocolumn,longbibliography]{revtex4-2}
\usepackage{graphicx}
\usepackage{bm,amsmath}
\usepackage{dcolumn}
\usepackage{amsfonts,amssymb}
\usepackage{chemfig}
\usepackage{subfigure}

\usepackage[colorlinks=true,linkcolor=blue,citecolor=blue,urlcolor=blue]{hyperref}

\begin{document}

\title{Single-hole spectral functions in one-dimensional quantum magnets with different ground states}
\author{Sibin Yang}
\affiliation{Department of Physics, Boston University, 590 Commonwealth Avenue, Boston, Massachusetts 02215, USA}

\author{Gabe Schumm}
\affiliation{Department of Physics, Boston University, 590 Commonwealth Avenue, Boston, Massachusetts 02215, USA}

\author{Bowen Zhao}
\affiliation{Department of Physics, Boston University, 590 Commonwealth Avenue, Boston, Massachusetts 02215, USA}

\author{Anders W. Sandvik}
\email{sandvik@buphy.bu.edu}
\affiliation{Department of Physics, Boston University, 590 Commonwealth Avenue, Boston, Massachusetts 02215, USA}

\date{\today}

\begin{abstract}
Recent advances in numerical analytic continuation with physics-motivated constraints now allow spectral functions
with sharp features, such as peaks and edges, to be extracted from imaginary-time correlation functions
computed by quantum Monte Carlo (QMC) simulations. Here, we test these approaches on various
one-dimensional $S=1/2$ spin systems with a single ejected fermion, i.e., extracting the single-hole
momentum ($k$)- and energy ($\omega$)-dependent spectral function $A(k,\omega)$. We compute the distance
($r$)- and imaginary time ($\tau$)-dependent Green's function $G(r,\tau)$ via Angelucci's canonical
transformation [Phys. Rev. B \textbf{51}, 11580 (1995)] of the fermionic Hamiltonian, implementing it for
stochastic series expansion QMC simulations. Our calculations of $A(k,\omega)$ focus on the different
characteristics of systems with spin-charge separation and those in which a spin polaron forms instead
because of effectively attractive interactions between the spin and the charge. Spin-charge separation is
well established in the conventional $t$-$J$ chain, which we confirm here as a demonstration of the method.
Turning on a multispin interaction $Q$ that eventually drives the system into a spontaneously dimerized
(valence-bond solid, VBS) state, we can observe the features of spin-charge separation until the VBS transition
takes place. While generally good agreement is found with the conventional analytical spin-charge separation
ansatz, we point out the formation of a gap between two holon bands that in the ansatz are degenerate
(crossing each other versus the momentum) at $k=0$ and $k=\pi$. Inside the VBS phase, effectively attractive
interactions may lead to the binding of the spinon and holon, for which we find evidence at large $Q/J$.
In the statically dimerized $t$-$J$ chain (i.e., with alternating strengths of the model parameters), we 
find equally spaced spin polaron bands corresponding to increasingly large bound states with two internal spin 
polaron modes---even and odd with respect to parton permutation. Our results overall demonstrate the power 
of modern analytic continuation tools in combination with large-scale quantum Monte Carlo simulations.
\end{abstract}
\maketitle

\section{Introduction}\label{sec:intro}

The single-hole spectral function is central to understanding the nature of low-energy excitations in strongly correlated electron
systems \cite{Hasegawa89,Dagotto90,Dagotto92,Poilblanc93,Poilblanc93_2,Preuss94,Beran96,Sorella96,Sorella96_2,Suzuura97,penc97,Sorella98,Martins00,Capponi01,Al-Hassanieh08,Maier08}. A prominent
example of exotic excitations (beyond those of Fermi liquid theory) arises in some one-dimensional (1D) antiferromagnets, where the single-hole dynamics
reveal fractionalized excitations---spinons (carrying spin $S=1/2$) and holons (carrying charge $C=1$)---that give rise to a continuum in the corresponding
spectral function accessible in angle-resolved photoemission spectroscopy (ARPES). This spin–charge separation contrasts with the spinon-holon bound
state, often called the spin polaron, which is a quasiparticle carrying both the spin and the charge that compensates for the ejected electron. Spin-charge
separation has been the subject of extensive theoretical investigation \cite{Lieb68,Voit95,Arikawa01,Arikawa04}, with substantial confirmation by numerical model studies
\cite{Kollath05,Kollath06,Poilblanc06,Smakov07,Wrzosek24} and experimental observations in quasi-1D materials
\cite{Kim96,Kim97,Fujisawa99,Recati03,Koitzsch06,Kim06,Vijayan20}. In addition to
their intrinsic importance in quantum many-body physics, work on spin-charge separation in 1D systems is also strongly motivated by possible analogies
in two-dimensional (2D) quantum magnets and their potential roles in high-$T_c$ superconductivity \cite{Bednorz86,Trugman88,Kane89,Martinez91,Poilblanc93,Poilblanc93_2}.

While the density-matrix renormalization group (DMRG) method \cite{White92,White93,Ostlund95,Schollwock05,Schollwock11,Al-Hassanieh08,Kollath05,Kollath06,Smakov07,Yang21,Yang22,Lin22,Drescher23} can now provide rather accurate results for various spectral functions of 1D systems
(and some 2D systems as well \cite{Lin22,Drescher23}), quantum Monte Carlo (QMC) simulations still have certain advantages, e.g., no increase in computational effort
with periodic boundary conditions (momentum conservation) and, typically, reliable convergence to the ground state (with decreasing temperature)
without potential trapping in metastable states. The challenge in dynamics studies with QMC is that a real-frequency spectral function $S(k,\omega)$
of an operator ${O}$ has to be constructed by numerical analytic continuation of the corresponding correlation function $G(\tau)=\langle O(\tau)O(0)\rangle$
in imaginary (Euclidean) time $\tau$. This inverse problem is ``ill-posed'' \cite{Jarrell96}, with the frequency resolution limited by the statistical
errors of the QMC data and, therefore, the outcome not being unique but to some extent depending on the method employed. With the recent development
of stochastic analytic continuation (SAC)~\cite{Sandvik98,Beach04} with physics-motivated constraints \cite{Sandvik16,Shao17,Shao23,Schumm24,Yang25,Schumm25}, it has
become possible to resolve the sharp spectral features expected at zero temperature, edges in particular. The frequency resolution can then exceed
that of DMRG calculations, where restrictions on time evolution \cite{White04} cause uncertainties in frequency space, similar to approaches
working directly in frequency space \cite{Jeckelmann02}. Open boundaries also limit the momentum resolution.

Constrained SAC has so far been employed mainly to study the dynamical spin structure factor of various spin models
\cite{Sandvik16,Shao17,Shao23,Schumm24,Yang25}. The approach was recently
also applied to the single-particle spectral function of the 2D half-filled Hubbard model, revealing novel features in the density of states
\cite{Schumm25}. This progress motivates us to revisit also the single-hole dynamics of spin systems. The hole dynamics is then governed
by the standard electron hopping terms, e.g., in the $t$-$J$ model with Hamiltonian
\begin{equation}\label{eq.tjhamiltonian}
H_{tJ} = -t \sum_{\langle ij\rangle,\sigma} 
      \left({c}_{i\sigma}^\dagger {c}_{j\sigma} + {c}_{j\sigma}^\dagger {c}_{i\sigma}\right) 
      - J \sum_{\langle ij\rangle} P_{ij},
\end{equation}
where ${c}_{i,\sigma}^\dagger$ (${c}_{i,\sigma}$) denotes the fermion creation (annihilation) operator with spin $\sigma=\uparrow,\downarrow$ on site $i$
(implicitly acting within the restricted Hilbert space without double occupancy) and $P_{ij}$ is the singlet projector on two sites defined as
\begin{equation}
P_{ij}=\tfrac{1}{4}n_in_j-\mathbf{S}_i \cdot \mathbf{S}_j.
\end{equation}
The sums in Eq.~(\ref{eq.tjhamiltonian}) typically run over nearest neighbors $\langle ij\rangle$, which will be the case considered here as well.

We consider several 1D systems, leaving the 2D case to separate publications \cite{Yang25new,Yang25new2}.
In addition to the conventional $t$-$J$ chain, we also consider its extension to a dimerized chain (with alternating strong and weak
couplings $t_i$, $J_i$). Moreover, we also investigate the uniform $t$–$J$–$Q$ chain,
with the $Q$ term defined as a product of three neighboring singlet projectors,
\begin{equation}\label{eq.tjqhamiltonian}
H_{tJQ} =  H_{tJ} - Q \sum_{\langle ijklmn\rangle} P_{ij} P_{kl} P_{mn},
\end{equation}
where $\langle ijklmn\rangle$ indicates six consecutive sites and the sum is over all such translated segments of sites, again with periodic boundary
conditions. Beyond a critical $Q/J$ ratio, the correlated singlet projection stabilizes a spontaneously dimerized doubly-degenerate
ground state (a valence-bond solid, VBS). A VBS phase also exists with only two projectors $P_{ij}P_{kl}$ \cite{Tang11,Tang13} in the $Q$ terms,
but here we use three of them in order to achieve a more
robust dimerization (maximized when $J=0$). The dimerization transition of the host spin chain is in the same universality class as the conventional
frustrated $J_1$-$J_2$ Heisenberg chain \cite{Majumdar69,Majumdar70,Okamoto92,Nomura94}, driven by a marginal operator as described by the level $k=1$
Wess-Zumino-Witten field theory \cite{Wess71,Witten83,Witten84}.

Carrying out QMC calculations of the imaginary-time correlators
\begin{equation}\label{gdef}
G_{ij}(\tau)=\langle {c}_{j\sigma}^\dagger(\tau) {c}_{i\sigma}(0)\rangle,
\end{equation}
the spectral function $A(q,\omega)$ is subsequently obtained by Fourier transformation and analytic continuation. We use the Stochastic Series Expansion
(SSE) QMC method \cite{Sandvik99,Sandvik10} supplemented by a trick previously developed and used with other QMC methods \cite{Angelucci95,Brunner00_1,Brunner00_2};
a short overview of the method and its adaptation to SSE will be provided below. While SSE and other QMC calculations are inhibited by the sign problem \cite{Loh90} for systems with finite hole density and hopping beyond nearest neighbors even in 1D systems, and even with only nearest-neighbor hopping at certain filling fractions, we here only consider nearest-neighbor hopping and there is no sign problem. But, regardless, it should be noted that the method used here only relies on importance sampling of the sign-free host system without dopants, and the injected hole exists only in a separate Hilbert space accessed for the ``measurements'' of the correlations. Thus there is no sign problem in higher dimensions either, even though there is some fraction of negative-amplitude measurements \cite{Brunner00_2,Yang25}.

One of the main goals of the present work is to test the ability of constrained SAC to reproduce known or expected behaviors in systems with
spin-charge separation and spin polaron formation. Beyond testing, the higher frequency resolution and the ability to resolve sharp edges and narrow
quasiparticle peaks also allow us to go beyond previous studies in characterizing the excitations and their spectral signatures.

To clarify the scope of the paper, we organize the work around three linked goals. First, we formulate and test the SSE estimator for the single-hole
Green's function against exact diagonalization on small systems. Second, we examine which SAC constraints are needed to resolve the sharp edges and
quasiparticle peaks expected in single-hole spectra. Third, we use these tools to distinguish spin-charge separated spectra from cases where the data
support spin-polaron formation. This organization separates the numerical-method tests from the physical conclusions drawn from the spectra.

The remainder of this paper is organized as follows: In Sec.~\ref{sec:methods}, we briefly outline the main steps for evaluating the single-hole
imaginary-time correlation function using SSE QMC simulations and introduce the basic principles of unconstrained SAC. The single-hole spectral function
$A(k,\omega)$ of the uniform $t$–$J$ and $t$–$J$–$Q$ chains is presented in Sec.~\ref{sec:1dtJQ}, where constrained sampling in SAC is also introduced
and results obtained with different constraints are compared and discussed. In Sec.~\ref{sec:dimerized}, we analyze multiband spin-polaron features
observed in the dimerized chain. Section~\ref{sec:summary} provides a summary of our findings and the advantages of constrained SAC.

\section{Methods}
\label{sec:methods}

In Sec.~\ref{subsec:itcf}, we explain the procedure for calculating the single-hole imaginary-time Green's function $G(i,j,\tau)$, adapting the procedures
of Refs.~\cite{Angelucci95,Brunner00_1,Brunner00_2} to the SSE representation of the imaginary-time continuum (instead of the time-discretized world-line
formalism used previously). We then perform a Fourier transform to obtain $G(k,\tau)$, which serves as the input for SAC after an additional processing
step involving constructing and diagonalizing the covariance matrix. In Sec.~\ref{subsec:sac}, we summarize the basic aspects of the SAC method without
sampling constraints. The constrained versions of SAC will be discussed in their respective contexts of different excitations with sharp features in
the frequency dependence.

\subsection{Single-hole Green's function}
\label{subsec:itcf}

For clarity, we first focus on the $t$–$J$ model, Eq.~(\ref{eq.tjhamiltonian}), and the almost trivial extension to the $t$–$J$–$Q$ model will be
explained later. The spin operators now have to be defined explicitly in terms of the fermionic operators:
\begin{eqnarray}\label{spinS}
  \mathbf{S}_i&=& (S^x_i,S^y_i,S^z_i) \\
   &=& \frac{1}{2}({c}_{i\uparrow}^\dagger {c}_{i\downarrow}+{c}_{i\downarrow}^\dagger {c}_{i\uparrow},
  i({c}_{i\downarrow}^\dagger {c}_{i\uparrow}-{c}_{i\uparrow}^\dagger {c}_{i\downarrow}),n_{i\uparrow}-n_{i\downarrow}). \nonumber
\end{eqnarray}
In order to simplify the Hamiltonian, we next perform Angelucci's canonical transformation \cite{Angelucci95},
\begin{equation}\label{cano_trans}
c^\dag_{i\uparrow}=\gamma_{i,+}f_i-\gamma_{i,-}f^\dag_i, \quad c^\dag_{i\downarrow}=\sigma_{i,-}(f_i+f_i^\dag),
\end{equation}
where $f_i$ are standard anticommuting spinless fermions, $\{f^\dag_i,f_j\}=\delta_{ij}$, and
\begin{subequations}
\begin{eqnarray}
\gamma_i^\pm&=&(1\pm\sigma_i^z)/2,\\
\sigma_i^\pm&=&(\sigma_i^x\pm i\sigma_i^y)/2,
\end{eqnarray}
\end{subequations}
where $\vec{\sigma}$ denotes the Pauli matrices, with $\sigma_{i}^{\pm}$ the corresponding spin raising and lowering operators.
Then, the Hamiltonian becomes
\begin{equation}\label{cano_transH}
H=t\sum_{\langle ij\rangle} \Lambda_{ij}\left(f_i^\dag f_j+f_j^\dag f_i \right)+\frac{J}{2}\sum_{\langle ij\rangle}\Delta_{ij}(\Lambda_{ij}-1),
\end{equation}
where
\begin{subequations}
\begin{eqnarray}
\Lambda_{ij}&=&(1+\vec{\sigma}_i\cdot\vec{\sigma}_j)/2\\
\Delta_{ij}&=&(1-n_i-n_j).
\end{eqnarray}
\end{subequations}
Note that the $J$ term is now expressed solely with the spin operators and the diagonal fermion number operators, $n_i \in \{0,1\}$, while the
kinetic $t$ term mixes spin interactions and fermion hopping. In the absence of holes, the Hamiltonian is still identical to that of the
$S=1/2$ antiferromagnetic Heisenberg model.

The fermion operators can be taken to represent degrees of freedom different from the spins but with an explicit constraint that is
conveniently taken into account by defining the single-site Hilbert space to consist of three states defined by
\begin{equation}\label{etabasis}
|\eta_{i}\rangle=|n_{i},z_{i}\rangle \in\left\{|0, \uparrow\rangle_{i},|0, \downarrow\rangle_{i},|1, \uparrow\rangle_{i}\right\},
\end{equation}
i.e., the hole can be regarded as replacing an $\uparrow$ spin. The ``missing state'' $|1, \downarrow\rangle_{i}$ represents a doubly occupied
site in the more general treatment of the Hubbard model \cite{Angelucci95}. It will also be useful to introduce a notation for the more
restricted Hilbert space without a hole
\begin{equation}\label{alphabasis}
|\alpha_{i}\rangle=|0,z_{i}\rangle \in\left\{|0, \uparrow\rangle_{i},|0, \downarrow\rangle_{i}\right \},
\end{equation}
which is just the standard $S=1/2$ spin space that will be used in the SSE sampling of the spin part of the Hamiltonian in the absence of
holes. The Hilbert space including a single hole in Eq.~(\ref{etabasis}) will here be accessed only when measuring the Green's function.

\begin{figure}[t]
\includegraphics[width=6cm]{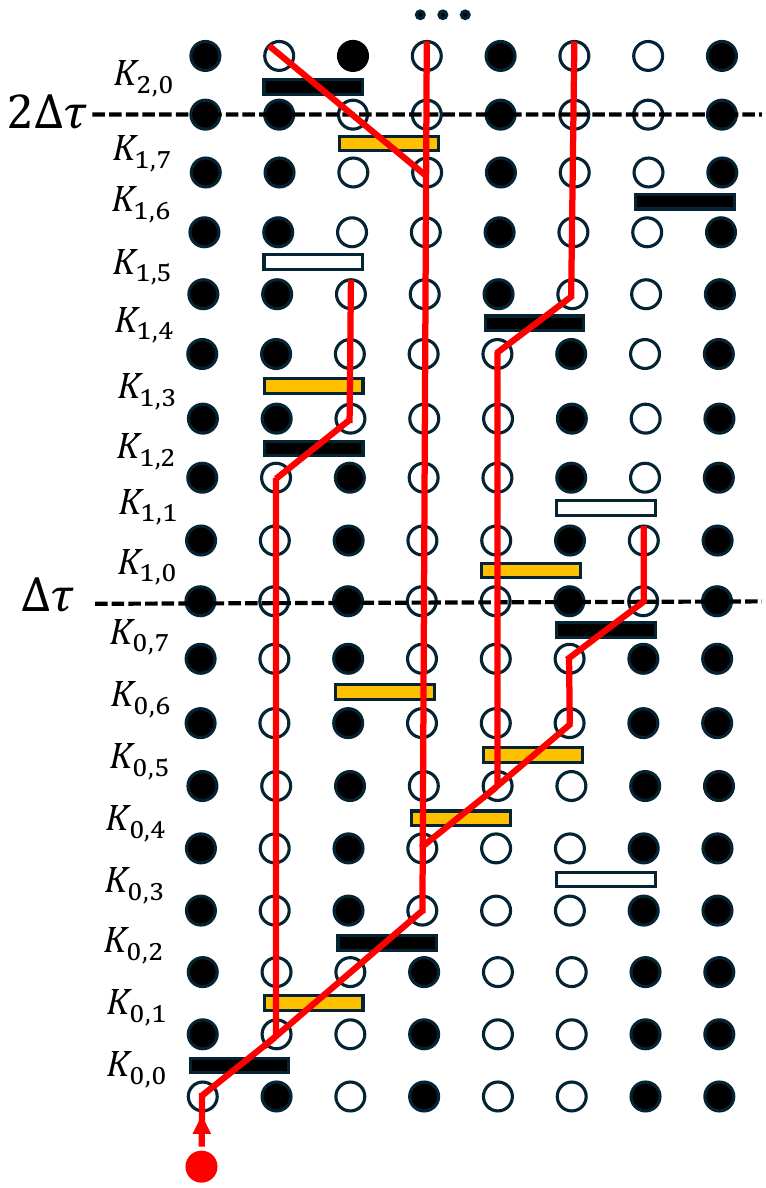}
\caption{Illustration of an SSE configuration of a system with eight spins and time slices including $M=8$ operators in Eq.~(\ref{ssez}). Spins $\uparrow$ and
$\downarrow$ are represented by open and solid circles, respectively, and the operators $K_{l,s}$ shown as bars are drawn from the sets defined in Eqs.~(\ref{hops}),
with the operator types corresponding to colors as follows: white for $a=1$, black for $a=2$, and yellow for $a=3$. No fill-in unit operators are 
included here but will appear in actual SSE configurations. The red dot represents an injected
hole replacing the $\uparrow$ spin at site $1$, and the red line illustrates the different paths this hole can take upon evolution in imaginary time.
Note that the hole in this representation can propagate only on sites with $\uparrow$ spins, reflecting the definition of the three-state basis,
Eq.~(\ref{etabasis}).}
\label{Fig.holemoves}
\end{figure} 

Three types of operators in the Hamiltonian can be identified that will be sampled in the SSE scheme:
\begin{subequations}
\begin{eqnarray}
-H_{1,ij}&=& J\Delta_{ij}(1-\sigma^z_i\sigma^z_j)/4,~~ \label{hop1} \\
-H_{2,ij}&=& -(\sigma^+_j\sigma^-_i +\sigma^-_j\sigma^+_i)\times \nonumber\\
  ~~~&&\left [J\Delta_{ij}/2+t({f}_{i}^\dagger {f}_{j} + {f}_{j}^\dagger {f}_{i})\right],~~\label{hop2}\\
-H_{3,ij}&=& -(t/2)(1+\sigma^z_i\sigma^z_j)({f}_{i}^\dagger {f}_{j} + {f}_{j}^\dagger {f}_{i}) + tI_{ij},~~ \label {hop3}.
\end{eqnarray}
\label{hops}
\end{subequations}
and the $t$-$J$ Hamiltonian is then
\begin{equation}
H = -\sum_{a=1}^3\sum_{\langle i,j\rangle} (-H_{a,ij}),
\end{equation}
where the first negative sign is canceled by the minus sign appearing in the density operator ${\rm e}^{-\beta H}$.
In Eq.~(\ref{hop3}), $I_{ij}$ is simply a unit operator associated with the sites $i,j$ whose role will become clear below.

To measure the Green's function at $L$ predefined time slices, which we here take equally spaced at separation $\Delta_\tau=\beta/L$, it is convenient to expand the individual slice operators ${\rm e}^{-\Delta_\tau H}$ instead of the single expansion of the full density operator \cite{Flynn26};
\begin{equation}\label{expslicing}
{\rm e}^{-\beta H} = \left ({\rm e}^{-\Delta_\tau H} \right )^L 
= \sum_{K}\left [\prod_{l=1}^{L} \frac{\Delta_\tau^{n_l}}{n_l!} \right ] \prod_{l=1}^{L}\prod_{s=1}^{n_l} K_{l,s},
\end{equation}  
where the operators $K_{l,i}$ are drawn from the set of operators defined in Eqs.~(\ref{hops}), i.e., $K_{l,s} \in \{-H_{a,ij}\}$. As usual \cite{Sandvik10},
the expansions are cut at some order $n_l=M$ beyond which the series is in practice completely converged. Unit operators $K_{l,s} \equiv I$ distinct
from the Hamiltonian operators are used to randomly augment the terms with $n_l < M$ to length $M$, with $n_l$ still counting the number of Hamiltonian
operators in the expansion at slice $l$. The combinatorial factor is then modified to
\begin{equation}
C(\{n_l\}) = \prod_{l=0}^{L-1} \Delta_\tau^{n_l}\frac{(M-n_l)!}{M!},
\end{equation}
and the partition function $Z={\rm Tr}\{{\rm e}^{-\beta H}\}$ is further sampled in a chosen basis, here the spin states
$|\alpha\rangle = \prod |\alpha_i\rangle$:
\begin{equation}\label{ssez}
Z = \sum_{\alpha,K} C(\{n_l\}) \left \langle \alpha \left | \prod_{l=0}^{L-1}\prod_{s=1}^{M} K_{l,s} \right | \alpha \right \rangle,
\end{equation}
where now $K_{l,s} \in \{-H_{a,ij},I\}$.
In practice $n_l < M$ is required to hold for all terms generated by choosing $M$ sufficiently large during the equilibration of the simulation,
so that the truncation does not introduce any error.

The expansion in Eq.~(\ref{ssez}) is positive definite for any antiferromagnetic spin model on a bipartite graph, on account of the trace in this case requiring
an even number of the operators $H_{2,ij}$ to be present in Eq.~(\ref{hop2}). Apart from the unit operators $H_{3,ij}$ in Eq.~(\ref{hop3}), which are
present even in the absence of holes, this is just the standard SSE formulation of the problem, and inclusion of the $H_{3,ij}$ operators in the sampling
is trivial. Fig.~\ref{Fig.holemoves} shows an illustration of an SSE configuration, also including an injected hole that we consider next. In this illustration
we have not indicated any fill-in unit operators in the truncated expansion, since they do not affect the hole motion. Actual SSE configurations with the cutoff
$M$ properly adjusted would have a significant fraction of unit operators (typically about 20-30\% \cite{Sandvik10}) in all the time slices.

Since no double-occupancy is allowed in the host spin model, the one-hole Green's function defined in Eq.~(\ref{gdef}) is expressed only with particle
destruction at a site $i$ at some time $\tau_0$, here taken as $\tau_0=0$ (later averaging over several slices for the injection), and creation at a site
$j$ at a later time $\tau_0+\tau$. Let us examine the action of the fermion destruction (hole creation) operators, the conjugates of Eq.~(\ref{cano_trans}),
on the single-site basis states in the notation of Eq.~(\ref{etabasis});
\begin{subequations}
\begin{eqnarray}
c_{i\uparrow}|0,\uparrow\rangle && =  f^\dagger_i |0,\uparrow\rangle = |1,\uparrow\rangle,\\
c_{i\downarrow}|0,\downarrow\rangle && = \sigma_i^+f^\dagger_i |0,\downarrow\rangle = |1,\uparrow\rangle,  \\
c_{i\uparrow}|0,\downarrow\rangle && = 0, \\
c_{i\downarrow}|0,\uparrow\rangle && = 0. 
\end{eqnarray}
\end{subequations}
Thus, the hole creation on an $\uparrow$ site is slightly simpler and we will consider only that case, knowing that destruction of a $\downarrow$
will lead to the same result by symmetry. Given the spin inversion symmetry of the models studied here, for the purpose of improved sampling statistics, 
hole creation on also $\downarrow$ sites can be considered by applying $c_{i\uparrow}$ to the inverted spin configuration in case of a $\downarrow$,
spin on $i$. With an $\uparrow$ spin at $i$ we have
\begin{align}\label{gtau1}
G(i,j,\tau) & =\langle c^\dag_{i\uparrow}(\tau)c_{j\uparrow}(0)\rangle=\langle f_{i}(\tau) f_{j}^{\dagger}(0)\rangle \notag\\
& =\frac{1}{Z} \operatorname{Tr}\left[e^{-(\beta-\tau) \mathcal{H}} f_{i} e^{-\tau \mathcal{H}} f_{j}^{\dagger}\right],
\end{align}
where the trace is over all states not containing a hole, i.e., the $\alpha$ basis defined in Eq.~(\ref{alphabasis}). Between the creation and destruction
operators, the evolved states contain a single hole and the basis states for a system of $N$ sites are products of a single-hole state $\eta_i=|1,\uparrow\rangle_i$
and $N-1$ spin states, $\prod |\alpha_j\rangle$, $j \not=i$. We refer to these states generically as $|\eta\rangle$.

Restricting the time $\tau$ to an integer $L_\tau$ of time slices, we proceed in analogy with the partition function in Eq.~(\ref{expslicing}),
writing
\begin{equation}
G(i,j,\tau)= \frac{1}{Z} \operatorname{Tr}\left[\prod_{l=L_{\tau}+1}^{L} e^{-\Delta \tau \mathcal{H}} f_{i}
\prod_{l=1}^{L_{\tau}} e^{-\Delta \tau \mathcal{H}} f_{j}^{\dagger}\right]. 
\end{equation}
and then inserting complete sets of states between all the exponential operators. Taking the trace and Taylor expanding all exponentials we have
\begin{widetext}
\begin{align}\label{gtaufull}
  G(i,j, \tau) = & \frac{1}{Z}
 \sum_{\alpha\eta}\langle\alpha_{0}| e^{-\Delta_\tau H}|\alpha_{L-1}\rangle \ldots\langle\alpha_{L_{\tau}+1}| e^{-\Delta_\tau H}|\alpha_{L_{\tau}}\rangle 
 \langle\alpha_{L_{\tau}}| f_{i} e^{-\Delta_\tau H}|\eta_{L_{\tau}-1}\rangle \ldots\langle\eta_{1}| e^{-\Delta_\tau H} f_{j}^{\dagger}|\alpha_{0}\rangle \notag\\
 = & \frac{1}{Z} \sum_{\alpha,K}C(\{n_{l}\})\langle\alpha_{0,0}| K_{L,M}|\alpha_{L-1,M}\rangle \cdots
 \langle\alpha_{0,2}|K_{1,2}|\alpha_{0,1}\rangle\langle\alpha_{0,1}|K_{1,1}|\alpha_{0,0}\rangle \times \notag \\
 &  ~~\sum_{\eta'}\frac{\langle\alpha_{L_{\tau},0}|f_{i}K_{L_{\tau},M}|\eta_{L_{\tau}-1,M-1}\rangle}{\langle\alpha_{L_{\tau},0}|K_{L_{\tau}, M}|\alpha_{L_{\tau}-1, M-1}\rangle} \cdots
   \frac{\langle\eta_{0,2}|K_{1,2}|\eta_{0,1}\rangle}{\langle\alpha_{0,2}|K_{1,2}|\alpha_{0,1}\rangle}
   \frac{\langle\eta_{0,1}|K_{1,1} f_{j}^{\dagger}|\alpha_{0,0}\rangle}{\langle\alpha_{0,1}|K_{1,1}|\alpha_{0,0}\rangle}   
   \notag \\
=& \sum_{\alpha,K} W(\alpha,K) G_{ij\tau}(\alpha,K).
\end{align}
\end{widetext}
Here $W(\alpha,K)$ is defined by the expression inside the summation on the
second line and is identical to the sampling weight of the partition function Eq.~(\ref{ssez}). The estimator
$G_{ij\tau}(\alpha,K)$ for the Green's function is the entire sum over states $|\eta'\rangle$ that represent the single-hole basis states inserted between
the time slices where a hole is present.
 The reason that this expression works is that the operator strings that contribute to the partition function of
the spin system are exactly the same ones that contribute when a single hole is present. The matrix elements are different however, as different parts of
the Hamiltonian operators in Eq.~(\ref{hops}) are active in propagating from one basis state to another basis state (with other parts annihilating the state).

With the two-body operators of the $t$-$J$ model, the bra and ket states appearing in nonvanishing matrix elements $\langle \alpha' |H_{a,ij}|\alpha\rangle$
and $\langle \eta' |H_{a,ij}|\eta\rangle$ differ at most at two sites $i,j$. Thus, the nontrivial elements can be expressed as $2\times 2$ matrices in
the relevant subspaces and can easily be read from Eq.~(\ref{hops}). For clarity, we here write these matrices together with the subspace states in question,
indicated on the right side of the respective matrices. Those in the sector with no holes are
\begin{subequations}
\begin{eqnarray}
  -H_{1,ij} &=& \frac{J}{2} \left [ \begin{array}{ll} 1 & 0 \\ 0 & 1 \end{array} \right ] 
\begin{array}{l} |0\uparrow, 0\downarrow\rangle  \\   |0\downarrow, 0\uparrow\rangle \end{array} \label{h1ijdef} \\   
  -H_{2,ij} &=& -\frac{J}{2} \left [ \begin{array}{ll} 0 & 1 \\ 1 & 0 \end{array} \right ]
\begin{array}{l} |0\uparrow, 0\downarrow\rangle  \\   |0\downarrow, 0\uparrow\rangle \end{array} \label{h2ijdef} \\
-H_{3,ij} &=& tI, \label{h3ijdef}
\end{eqnarray}
\label{haijdefs}
\end{subequations}
where the identity operator on the last line is defined in the full 2-spin space $|\sigma^z_i,\sigma^z_j\rangle$. In the one-hole sector 
\begin{subequations}
\begin{eqnarray}
  -H_{1,ij} &=& 0 \label{hh1ijdef} \\
  -H_{2,ij} &=& -t\left [ \begin{array}{ll} 0 & 1 \\ 1 & 0 \end{array} \right ]
\begin{array}{l} |1\uparrow,0\downarrow\rangle  \\   |0\downarrow,1\uparrow\rangle \end{array} \label{hh2ijdef} \\  
  -H_{3,ij} &=& t \left [ \begin{array}{ll} +1 & -1 \\ -1 & +1 \end{array} \right ]
  \begin{array}{l} |0\uparrow,1\uparrow\rangle  \\   |1\uparrow,0\uparrow\rangle \end{array} ~~{\rm or} \label{hh3ijdef} \\
   -H_{3,ij} &=& t \left [ \begin{array}{ll} 1 & 0 \\ 0 & 1 \end{array} \right ]
  \begin{array}{l} |1\uparrow,0\downarrow\rangle  \\ |0\downarrow,1\uparrow\rangle \end{array}. \label{hh4ijdef}
\end{eqnarray}
\label{hhaijdefs}
\end{subequations}
Here $0$ in Eq.~(\ref{hh1ijdef}) implies that this operator will destroy any state with a hole; thus, a hole propagation path will terminate when an
$H_{1,ij}$ operator (corresponding to a diagonal spin interaction) is encountered. Another key aspect of the hole propagation is that $H_{3,ij}$ when both
spins are $\uparrow$, as in Eq.~(\ref{hh3ijdef}), leads to branching, with a path splitting into one branch continuing to evolve on the same site while the
other branch contains the hole jumping to the second site of the operator. This operator can also appear between two antiparallel spins, in which case the hole can only
continue to propagate on the $\uparrow$ site, as indicated by the second matrix for this case in Eq.~(\ref{hh4ijdef}). The case of the hole entering this operator
at the $\downarrow$ spin is excluded from the outset (and would lead to zero). The initial part of a branching hole path is shown in Fig.~\ref{Fig.holemoves}.

The contributions to the Green's function in Eq.~(\ref{gtaufull}) are products of matrix element ratios $R_{a,ij}$, $a=1,2,3$, which from
Eqs.~(\ref{haijdefs}) and (\ref{hhaijdefs}) are
\begin{subequations}
\begin{eqnarray}
  R_{1,ij} &=& 0 \label{rat1} \\
  R_{2,ij} &=& \frac{2t}{J}\left [ \begin{array}{ll} 0 & 1 \\ 1 & 0 \end{array} \right ]
\begin{array}{l} |1\uparrow,0\downarrow\rangle  \\   |0\downarrow,1\uparrow\rangle \end{array} \label{rat2} \\  
  R_{3,ij} &=& \left [ \begin{array}{ll} +1 & -1 \\ -1 & +1 \end{array} \right ]
  \begin{array}{l} |0\uparrow,1\uparrow\rangle  \\   |1\uparrow,0\uparrow\rangle \end{array} ~~{\rm or} \label{rat3} \\
   R_{3,ij} &=& \left [ \begin{array}{ll} 1 & 0 \\ 0 & 1 \end{array} \right ]
  \begin{array}{l} |1\uparrow,0\downarrow\rangle  \\ |0\downarrow,1\uparrow\rangle \end{array}. \label{rat4}   
\end{eqnarray}
\label{ratios}
\end{subequations}
With a hole injected into an SSE configuration at a specific site $i$ at an allowed $\tau$ point (at which the spin is in the $\uparrow$ state),
the contributions from all possible paths can be summed up in the following way: A vector ${v}$ with $N$ elements (the number of spins),
is initialized with all zeros, except for a $1$ at the position $i$ corresponding to the $\uparrow$ site into which the hole is injected
into the state $|\alpha_{0,0}\rangle$ that is stored. The SSE operator string is then traversed, updating the stored spin state according to the
operators encountered sequentially in Eq.~(\ref{gtaufull}). For each Hamiltonian operator (disregarding the trivial unit operators
used to augment the operator sequence), the vector ${v}$ is multiplied by the corresponding matrix $R_{a,ij}\otimes I_{ij}$, where $I_{ij}$ formally
indicates the part of the $N\times N$ matrix comprising only $N-2$ diagonal unit elements. For each entire time slice traversed, the vector elements ${v}(j)$
represent the estimators of $G(i,j,\tau)$ for all $j$ when the starting position is $i$. Arranging the vectors for all allowable injection sites as
the columns of an $N/2 \times N$ matrix, with the $i$th column initialized with a $1$ in the element corresponding to the location of the $i$th $\uparrow$
spin as above (note that the number of $\uparrow$ sites is always $N/2$ when the temperature is low enough for only the singlet ground state of the spin
system to be sampled), the estimators for
all $G(i,j,\tau)$ are generated from a single scan of the SSE operators. The computational effort then scales as $N^2\tau$, with $N\tau$ coming from
the scaling $N\beta$ of the total number of operators in the SSE expansion. The scaling is the same as in the previous implementations within the
discrete (Trotter approximated) world-line method \cite{Brunner00_1,Brunner00_2} but in the SSE approach there are no discretization errors. The results
can also be averaged over injections at different time slices, which leads to another factor $\beta$ in the formal scaling but also provides additional
statistics if the injection slices are sufficiently far apart.

Moving on to the $t$-$J$-$Q$ model, SSE simulations of various $J$-$Q$ models require only minor modifications of the Heisenberg ($J$ model)
algorithm and have been described in detail in previous works \cite{Sandvik10}. With $Q$ terms defined as products of the singlet projectors
in Eq.~(\ref{eq.tjqhamiltonian}), they can be treated the same way as the $J$-operators. The hopping terms still need to be combined only with the $J$-operators
as in Eqs.~(\ref{hops}), and thus no holes are allowed on the additional operators corresponding to the $Q$ terms. The modification of the estimator
for the Green's function is then trivial---any hole branch entering a $Q$ operator is terminated, with a corresponding $0$ in the matrix element
ratio as in Eq.~(\ref{rat1}). Another option is to divide the $t$ terms also over the $Q$ interactions. While we have not made extensive comparisons, limited tests
indicate that it is somewhat more efficient to combine the $t$-terms only with the $J$-interactions, unless $J/Q$ is very small.

\begin{figure}
\includegraphics[width=8.0cm]{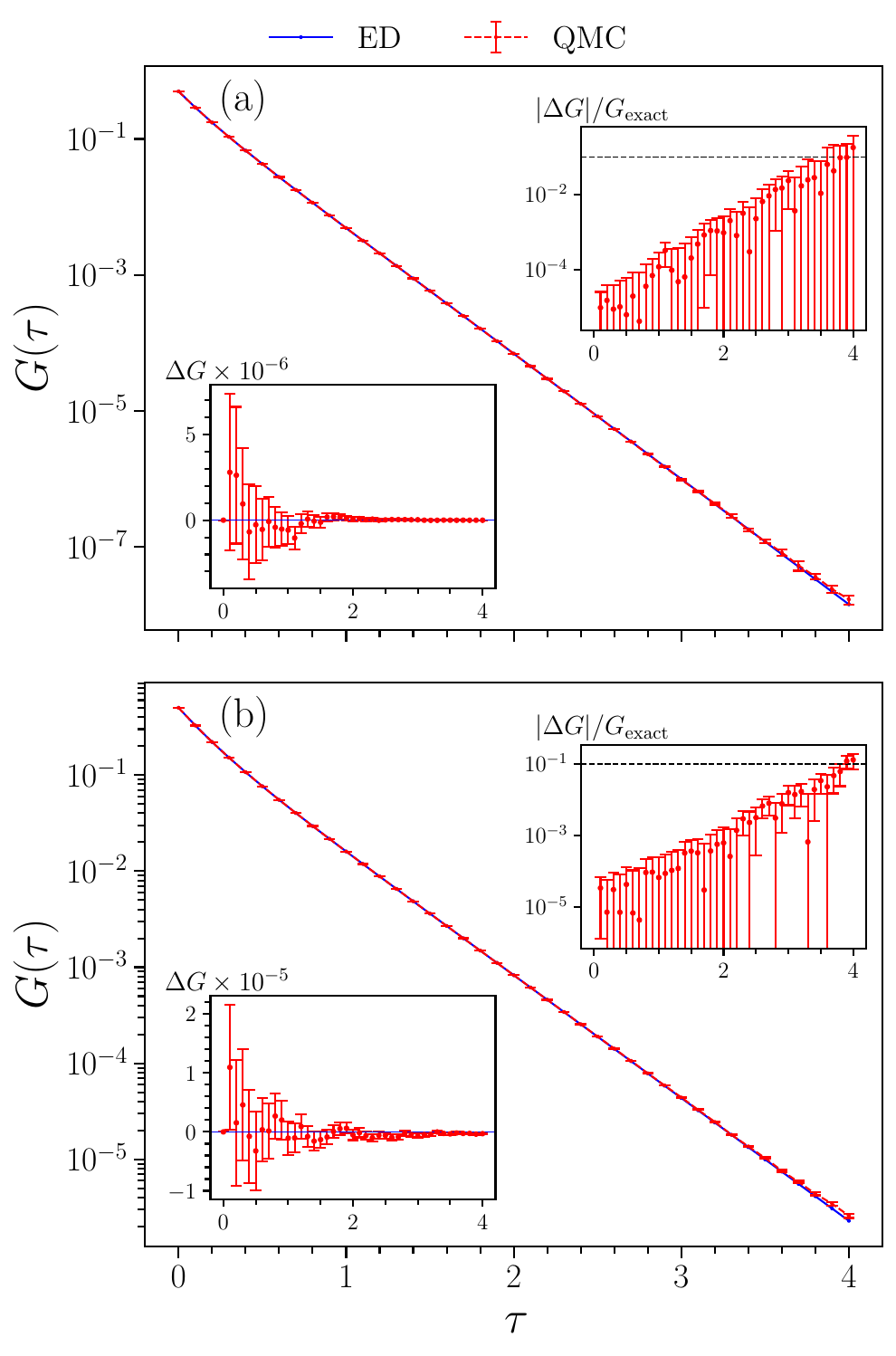}
\caption{Comparisons of the exact $k=\pi/2$ Green's function (blue curve, from full diagonalization of $H$) of $N=8$ chains and corresponding SSE results
(red points with error bars) for
(a) the standard $t$-$J$ chain at $t=1$, $J=4$ and (b) The  $t$-$J$-$Q$ chain with $t=1$, $J=2$, and $Q=0.5$. The left and right insets show, respectively,
the absolute and relative difference between the SSE data and the exact results. In the analytic continuation we use only data points for which the
statistical error (one standard deviation of the mean) is less than $0.1$, which corresponds to a similar relative error with respect to the exact
value, indicated here with dashed lines. Note that the log scale in the insets showing $|\Delta G|/G_{\rm exact}$ does not visually reflect well the
size of the lower part of the error bar (one standard deviation of the mean). In (a) the
deviation from the exact value at $\tau=4$ is $1.0$ standard deviation, while in (b) it is $2.1$ standard deviations. The deviations overall are
statistically likely.}
\label{Fig.EDcomps}
\end{figure} 

In practice, the relative statistical errors of $G(k,\tau)$ (with $k$ denoting the dimensionless momentum after Fourier transforming SSE results for
all site pairs $i,j$) grow with increasing $\tau$ and the calculation is therefore restricted to a number
of $\tau$ values below a cutoff $\tau_{\rm max}$. In the work presented here, we typically take $t=1$ as the energy scale, set the time slice to $\Delta_\tau=0.1$
and generate data up to $\tau=4$ or $\tau=8$. We also average over imaginary time, stepping over a number of time slices corresponding to the
maximum $\tau$ value used. Aiming for the ground state, we set the inverse temperature to a multiple of the system size $N$ (typically $\beta=2N$),
sufficiently large for no statistically significant finite-temperature effects to remain up to $\tau_{\rm max}$. To test the code, we have compared the results
with exact diagonalization for small systems; two illustrative examples of the Green's function $G(k,\tau)$ in momentum space are shown in Fig.~\ref{Fig.EDcomps}.
These ED comparisons are used as controlled benchmarks of the estimator and implementation.
The large-system spectral conclusions below rely on QMC data for longer chains combined with SAC, while the ED results serve to verify the imaginary-time
Green's functions on systems where exact results are available.
Here it is interesting to note that the magnitude of the statistical errors actually decreases with increasing $\tau$, reflecting the improved statistics
as the hole path branches out and eventually terminates---the averages at large $\tau$ are then over small values in many elements in the vector $v$
representing the propagated hole amplitudes, instead of a small number of larger values at small $\tau$. The relative statistical errors of the
asymptotically exponentially decaying $G(\tau)$ still increase with $\tau$.

\subsection{Spectral function}
\label{subsec:sac}

Here we consider the $T=0$ spectral function of the single-hole operator ${O}={c}_{{k},\uparrow}$, 
which in the Lehman representation is given by
\begin{equation} {\label{eq.singleholet0}}
A({k},\omega)=\pi\sum_{m} |\langle m,N-1|{c}_{{k},\uparrow}|0,N\rangle|^2\delta(\omega-\epsilon_{m}).
\end{equation}
Here, $|0,N\rangle$ denotes the ground state of the spin model (corresponding to a half-filled Hubbard model with large repulsion $U$) with $N$ spins
and ground state energy $E_0$ (and the momentum of this state is $k=0$ for a chain of length $L=4n$ for integer $n$),
while $|m,N-1\rangle$ are the states with a hole and energy $E_m$ (all of which
have momentum $-{k}$), and $\epsilon_m=E_m-E_0$ is the excitation energy. The corresponding correlation function at momentum ${k}$ and imaginary
time $\tau$
\begin{equation} {\label{eq.gtau}}
 G({k},\tau)=\langle {c}^\dagger_{{k},\uparrow}(\tau) {c}_{{k},\uparrow}(0)\rangle,
\end{equation}
is obtained by Fourier transforming the real-space correlators discussed above in Sec.~\ref{subsec:itcf}. The relation between $G({k},\tau)$ and
the real-frequency spectral function $A({k},\omega)$ at $T=0$ is
\begin{equation}\label{eq.Gtausw}
G({k},\tau)=\frac{1}{\pi}\int_{-\infty}^{+\infty} d\omega A({k},\omega) e^{-\tau \omega},
\end{equation}
which we invert using the SAC method.

\begin{figure}[t]
\includegraphics[width=8.3cm]{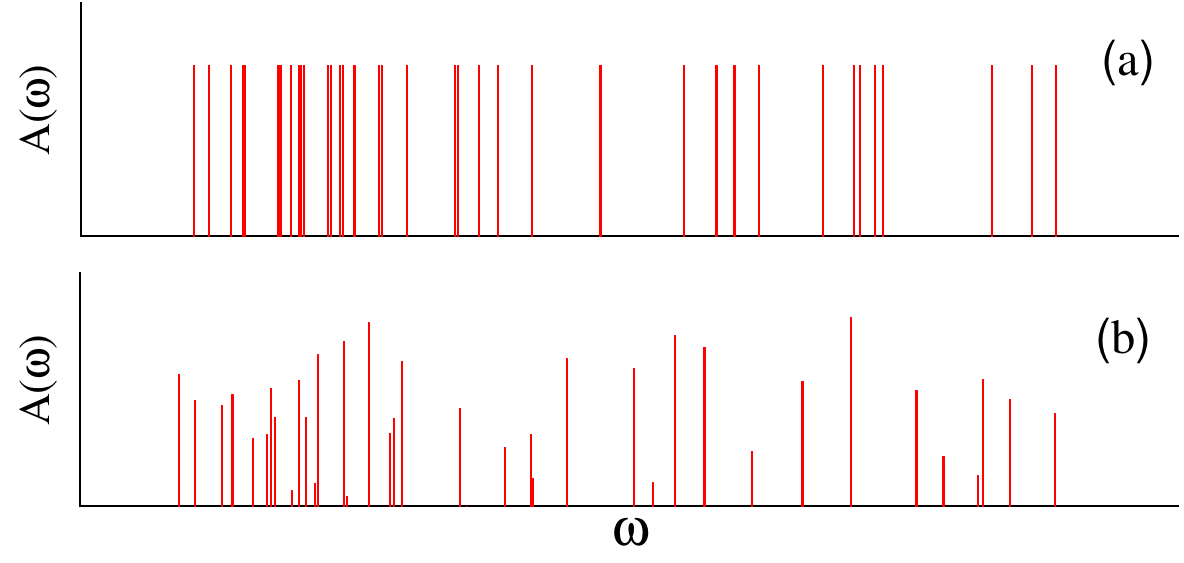}
\caption{Two parameterizations of the spectral function $A(\omega)$ used in unconstrained SAC with a large number $N_\omega$ (typically thousands)
of $\delta$-functions in the frequency continuum (in practice on a grid with very small spacing; $\Delta_\omega=10^{-5}$ or smaller). In (a), each
$\delta$-function carries the same fixed weight $N^{-1}_\omega$ of the normalized spectrum and only the locations $\omega_i$ are sampled, while in (b) also
the amplitudes are sampled, with the total spectral weight conserved.}
\label{Fig.sacpara1}
\end{figure} 

We summarize only the generic SAC ingredients needed to define the notation used below. In the unconstrained form, a large number $N_\omega$ of
$\delta$-functions is sampled and
(suppressing the $k$ label here)
\begin{equation}\label{eq.somegadeltas}
A(\omega) = \sum_{i=1}^{N_\omega} a_i \delta(\omega-\omega_i),
\end{equation}
with $N_\omega$ typically of order $10^3$ or larger. The frequencies $\omega_i$ are effectively continuous, implemented in practice on a very fine
grid between bounds where the spectral weight is negligible; this allows the kernel in Eq.~(\ref{eq.Gtausw}) to be precomputed. Fig.~\ref{Fig.sacpara1}
shows the two unconstrained parametrizations used as reference cases: fixed equal amplitudes \cite{Qin17} and variable amplitudes \cite{Beach04}.
Their different configurational entropies can lead to different artificial broadening or secondary features \cite{Shao23,Ghanem23}, which motivates
the constrained parametrizations introduced in the results sections.

The spectrum is sampled with the Boltzmann-like probability distribution
\begin{equation}\label{P(S)}
 P(A|\bar G) \propto \text{exp}\left (-\frac{\chi^2}{2\Theta}\right ),
\end{equation}
where the fictitious temperature $\Theta$ controls the compromise between fitting the QMC data and avoiding over-fitting to statistical noise. We use
the standard criterion \cite{Qin17,Shao23,Schumm24}
\begin{equation}\label{eq.criterion}
\langle\chi^2(\Theta)\rangle=\chi^2_\text{min}+a\sqrt{2\chi^2_\text{min}}.
\end{equation}
Here $a$ is a constant of order unity and $\chi^2$ is the goodness of fit to the QMC data defined with the full covariance matrix $C$,
\begin{equation}\label{eq.chi2cov}
\chi^2=\sum_i \sum_j (G_i-\bar G_i)[C^{-1}]_{ij} (G_j-\bar G_j),
\end{equation}
where $G_i\equiv G(\tau_i)$ is obtained from $A(\omega)$ using Eq.~(\ref{eq.Gtausw}) and $\bar G_i$ is the corresponding QMC mean value. The covariance
matrix is computed from binned QMC averages $G^b_i$, where the bin index $b \in \{1,\ldots,N_B\}$;
\begin{equation}
 C_{ij}=\frac{1}{N_B(N_B-1)}\sum_{b=1}^{N_B} (G^b_i-\bar G_i)(G^b_j-\bar G_j).
\end{equation}
with $N_B$ much larger than the number of $\tau$ points $N_\tau$ \cite{Jarrell96}. Further technical details and recent SAC developments are discussed
in Refs.~\cite{Shao23,Schumm24,Yang25,Schumm25}. The methodological elements emphasized below are the application of lower-edge and double-edge
constraints to single-hole spectra, and the use of quasiparticle plus second-gap parametrizations to separate isolated spin-polaron features from
continua.

\section{\texorpdfstring{Uniform $\mathit{t}$-$J$ and $\mathit{t}$-$J$-$Q$ Chains}{Uniform t-J and t-J-Q Chains}}
\label{sec:1dtJQ}

According to the spin-charge separation ansatz \cite{Suzuura97}, excitations formed by hole injection by the operator $c_{k,\sigma}$ are fractionalized
in the $t$-$J$ chain, thus leading to a continuum of spinon and holon contributions to $A(k,\omega)$. The spectral bounds are expected to be manifested
as edges with power-law divergences at the lowest and highest combined spinon and holon energies. Here we study the standard $t$-$J$ chain as well
as the $t$-$J$-$Q$ chain, in the latter first with $Q/J$ tuned to the dimerization transition point $(Q/J)_c$ and then inside the VBS phase for
$Q/J > (Q/J)_c$. At $(Q/J)_c$, the logarithmic corrections associated with the marginally irrelevant operator in the host Heisenberg chain ($t=0$
and no holes) vanish, changing signs and becoming marginally relevant in the VBS phase \cite{Affleck87,Affleck88}. While it is not clear how the logarithmic
corrections affect the single-hole spectral function, given that the spin-charge separation ansatz is only an approximation (a mean-field solution of the
Hamiltonian written with ``slave bosons'') \cite{Suzuura97,Brunner00_1} and does not account for subtleties such as logarithmic corrections. It is nevertheless interesting
to study the $t$-$J$-$Q$ chain at the special point $(Q/J)_c$.

At the mean-field level, the spinon and the holon propagate with independent momenta $q_s$ and $q_h$, respectively, with dispersion relations
\begin{subequations}
\begin{eqnarray}
  \epsilon_s & = & -J_s\cos(q_s), \label{disp_s}\\
  \epsilon_h & = & -t_h\cos(q_h). \label{disp_h}
\end{eqnarray}
\label{disp}
\end{subequations}
When the hole operator $c_{k}$ acts on the ground state of the spin chain, a holon
and an anti-spinon are created. Thus, the total momentum is $k=q_h-q_s$ and the energy is $E_k=\epsilon_h-\epsilon_s$. The parameters $t_h$ and
$J_s$ are equal to $2t$ and $J$ in the mean-field Hamiltonian according to Ref.~\cite{Brunner00_1}, but instead of the bare values they can also be taken as phenomenological parameters.
In the case of the $t$-$J$-$Q$ chain, we have not derived the mean-field expressions for $t_s$ and $J_s$ but simply take them as fitting parameters.
In all cases, we also have to add a constant to the holon dispersion, corresponding to a chemical potential, $\epsilon_h \to \epsilon_h + \mu$, where
we also treat $\mu$ as a fitting parameter.

While only a single hole is injected into the host spin system and only one holon is created, the resulting fermionic anti-spinon is taken from a Fermi sea of
already existing spinons. Furthermore, a so-called phase-string effect modifies the spectral weight density, leading to singularities in $A(k,\omega)$ beyond a
sharp edge with divergent spectral weight density that is present already with the bare spinon and holon. Thus, the behavior is quite complex even at the mean-field level.
We refer to the literature \cite{Suzuura97,Brunner00_1} for details and only quote results required to motivate a modified SAC parametrization used below.

As a consequence of the restriction of the spinon to its Fermi sea, there is a special momentum $k_0$, with $\cos(k_0)=J_s/t_h$, such that the lower edge
of the compact support of $A(k,\omega)$ is given by the function $-F_k$ for $k<k_0$ and the upper edge is at $F_k$ for $k_0 < k < \pi$, where
\begin{equation}
F_k=\sqrt{J_s^2+t_h^2-2t_hJ_s\cos(k)}.
\end{equation}
Because of the phase-string effect, these edges are associated with power-law singularities. The remaining edges of the compact support outside the regions
defined by $k_0$ above are given by $E_k = \pm t_h\sin(k)$, arising from holons with momentum $q_h = k+q_s$, $q_s=\pm \pi/2$ (i.e., $\epsilon_s=0$), and these
are also preceded by power-law divergent spectral weight. Thus, for all $k$, there are altogether three peaks versus $\omega$, though close to
$k_0$ the third peak will be very close to $E_k$ and, therefore, very hard to separate as an individual feature in numerical studies.

\subsection{Critical phase and dimerization point}
\label{sec:tjq_1}

\begin{figure}
\includegraphics[width=6.5cm]{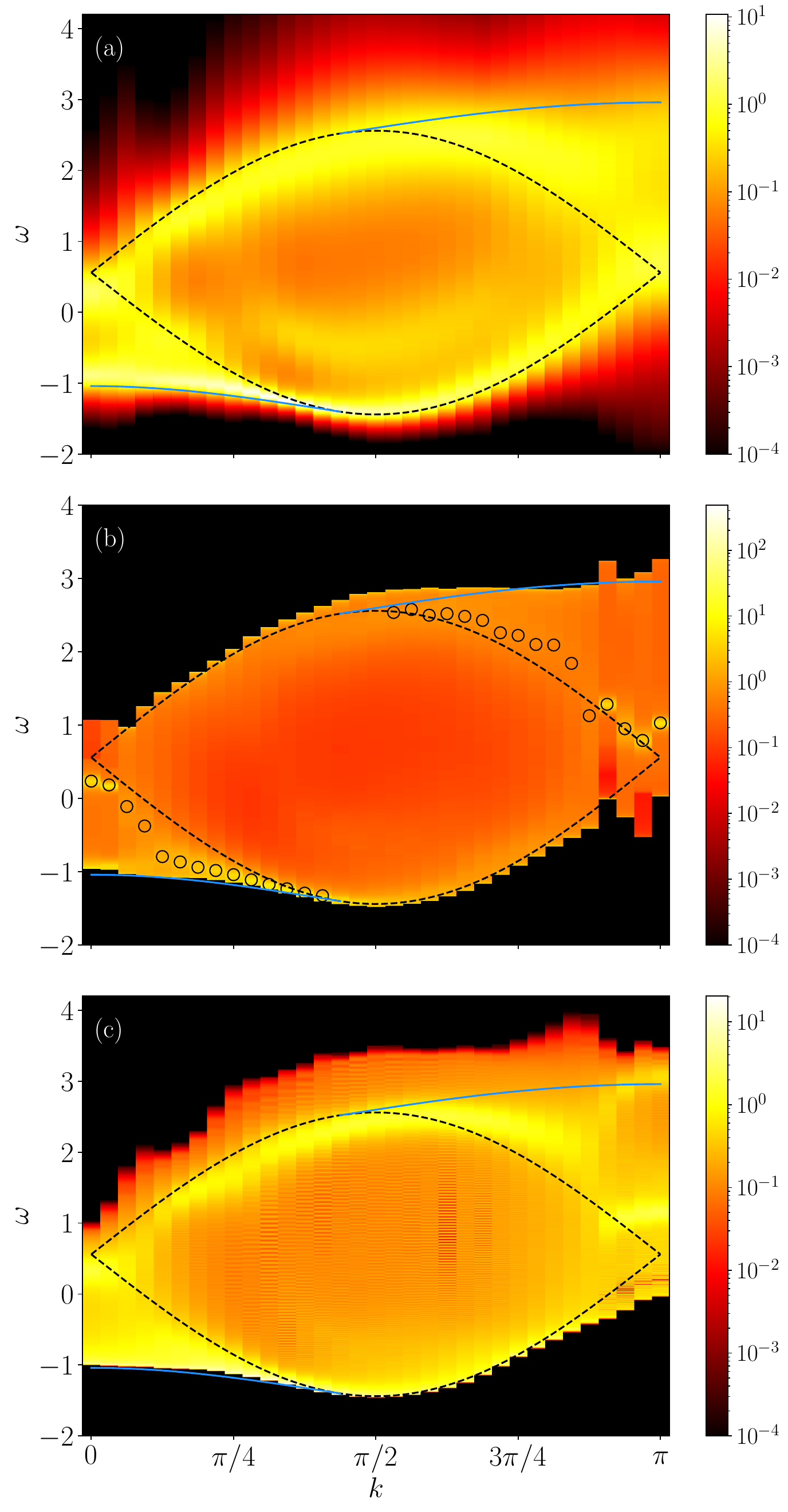}
\caption{Spectral function $A(k,\omega)$ of the $L=64$ $t$-$J$ chain with $t=1$ and $J=0.4$ obtained by SAC with one unconstrained and
two constrained $\delta$-function parametrizations. (a) with unconstrained variable amplitudes as in Fig.~\ref{Fig.sacpara1}(b). 
(b) with a double edge representing the spectral bounds and unconstrained contributions between the edges, combining Figs.~\ref{Fig.edgecontpara}(a),
\ref{Fig.edgecontpara}(b), and \ref{Fig.edgecontpara}(c). In (c), the single-edge constraint illustrated in Fig.~\ref{Fig.edgecontpara}(a) has been combined with 
unconstrained $\delta$-functions above the edge as in Fig.~\ref{Fig.edgecontpara}(b). The black dashed lines show the upper and lower holon branch (on which the spinon
has momentum $k=\pi/2$ and energy $0$) according to the spin-charge separation ansatz. The blue solid lines show the lowest and highest energies
of the combined spinon and holon excitations when not coinciding with the holon branches. These predictions are based on the dispersion relations
in Eqs.~(\ref{disp}) with $J_s=J$ and $t_h=2t$. The circles in (b) show the locations of local maximums inside the continuum that we identify
with the holon branches.}
\label{Fig.t1J04logcombine}
\end{figure} 

\begin{figure*}
\includegraphics[width=14cm]{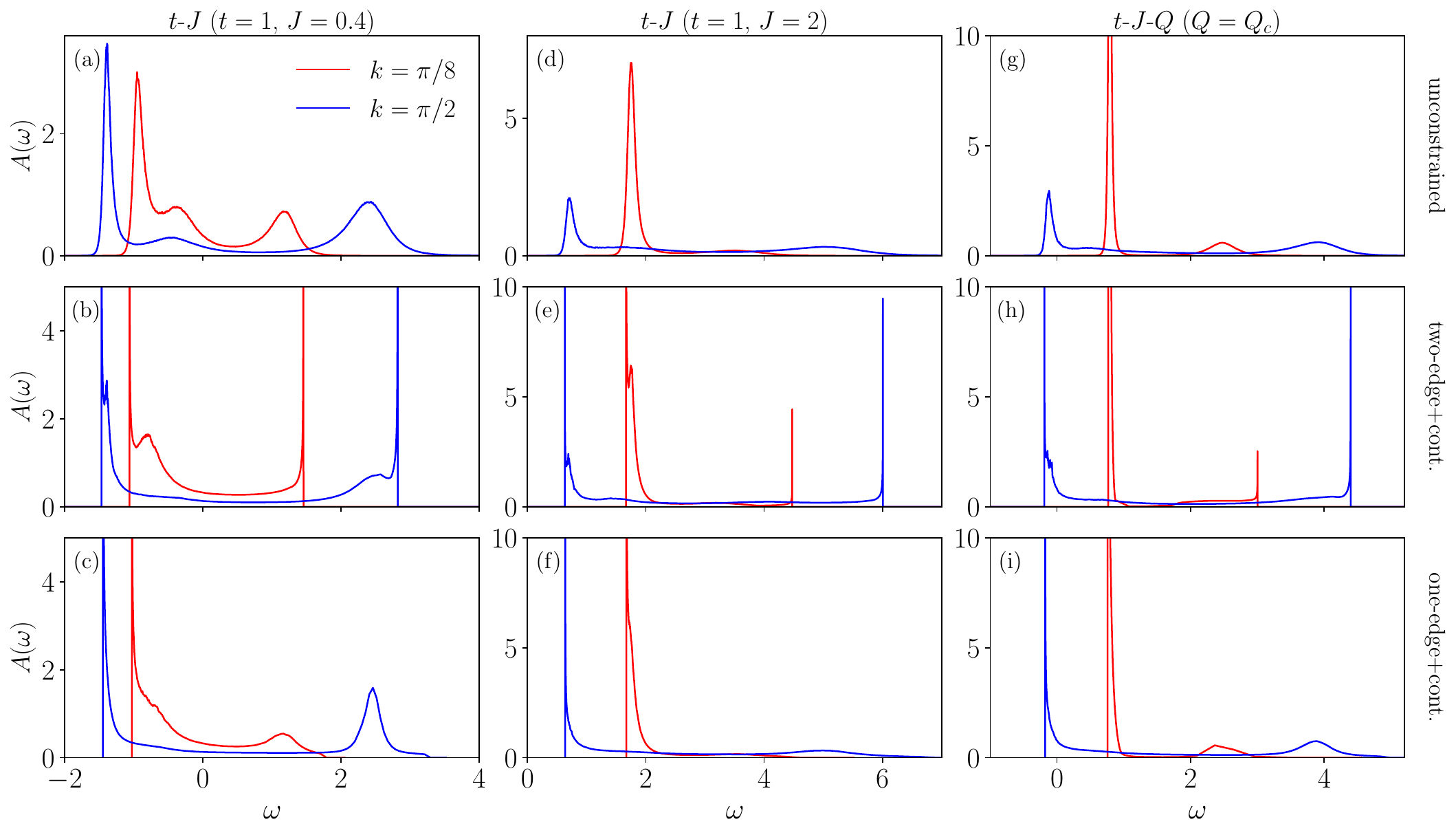}
\caption{Spectral profiles for fixed $k=\pi/8$ (red curves) and $\pi/2$ (blue curves) for three different parameter sets of the $t$-$J$ ($Q=0$) and $t$-$J$-$Q$
models, as indicated on top of the respective columns. In each case, three different SAC parametrizations were used; unconstrained in (a),(d),(g), double-edge in (b),(e),(h) and lower edge only in (c),(f),(i). Parameter sets correspond to the heat-map representation of the entire spectral function in
Figs.~\ref{Fig.t1J04logcombine}, \ref{Fig.t1J2combine}, and \ref{Fig.t1J1Q01645logmorecombine}.}
\label{Fig.sw_compare_all}
\end{figure*} 

We first consider the $t$-$J$ chain with $t=1$ and $J=0.4$, which is one of the cases used in previous QMC calculations with analytic continuation
by the maximum-entropy method by Brunner et al.~\cite{Brunner00_1} for chain lengths up to $L=128$. Recently, Wrzosek et al.~presented exact diagonalization
(ED) results for the same model parameters and a system of size $L=28$ \cite{Wrzosek24}, larger than in previous (ED) studies. All results here will be for $L=64$,
even though much larger systems can be accessed with our methods. Based on tests, we do not see significant finite-size changes for $L>64$, and since high-precision
$G(k,\tau)$ results are required for reliable analytic continuation it is better to use moderate system sizes for which smaller statistical errors can be achieved 
at reasonable computational cost. Fig.~\ref{Fig.t1J04logcombine} shows our analytically continued results based on three different SAC parametrizations, showing the
complete spectral function for all $0 \le k \le\pi$. The logarithmic scale in the heat map spans values down to $10^{-4}$, such that black regions correspond to 
negligible or vanishing spectral weight. To complement the heat map representation of $A(k,\omega)$, in Figs.~\ref{Fig.sw_compare_all}(a)-(c) we show examples of
spectral profiles graphed on linear scale for two values of $k$.

Figs.~\ref{Fig.t1J04logcombine}(a) and  Fig.~\ref{Fig.sw_compare_all}(a)
show results of unconstrained sampling with both frequencies and amplitudes updated, corresponding to Fig.~\ref{Fig.sacpara1}.
The expected singularities discussed above are marked, here with $t_h=2t$ and $J_s=J$, which represent very close to optimal parameters for this rather small value
of $J/t$ (where the ansatz is expected to work the best). The bright bands in the SAC spectra coincide very closely with the predicted spectral bounds, and there
is relatively little spectral weight above the predicted upper bound. There is elevated spectral weight around the predicted holon singularities at
$\omega_k = \pm t_h\sin(k) + \mu$ (with $\mu$ adjusted for best fit)
also when these singularities are inside the continuum. Overall, the results look rather similar to the exact diagonalization
results for $L=28$ \cite{Wrzosek24}, where additional
``mini bands'' are seen inside the continuum because of the small number of $\delta$ functions making up the exact spectral
function, Eq.~(\ref{eq.singleholet0}), for a small system. The SAC method cannot resolve such fine structure, which is anyway expected to smooth into a true
continuum in the thermodynamic limit. With unconstrained sampling, the expected pure holon singularity for $k < k_0$ (where the spinon momentum $k_s=\pi/2$ and,
therefore, $\epsilon_s=0$) is only marginally resolved as a broad peak, similar to previous results obtained with the maximum-entropy method \cite{Brunner00_1}.

In Fig.~\ref{Fig.t1J04logcombine}(a), there is also a weaker broad band close to the lower holon edge. It is known that numerical analytic continuation,
with SAC or other related techniques, e.g., the maximum-entropy method \cite{Jarrell96}, is prone to producing artificial peaks in spectral functions
(versus $\omega$ for fixed $k$) with QMC imaginary-time data $G(k,\tau)$ of typical statistical quality. It has been shown that such peaks often are secondary,
spurious features resulting from excessive broadening of peaks at lower energy \cite{Shao23}. In the present case, the exact spectrum should have a sharp lower
edge but it is broadened significantly by unconstrained SAC (as with any conventional analytic continuation method); in particular, spectral weight is present below the edge.
The wrongly distributed weight leads to $G(k,\tau)$ decaying slower than it should for large $\tau$, and to partially compensate (in order for the
sampled spectrum to fit the data) the peak location is pushed slightly up in frequency. A secondary peak typically also forms, which  can be regarded
as a secondary compensatory effect---when the main peak has been pushed up in frequency, a minimum forms on its right side to balance the mean spectral
density in this $\omega$ range. Thus, the U-shaped band between the holon bounds is likely a spurious feature, which we demonstrate explicitly next.

\begin{figure}
\includegraphics[width=8.3cm]{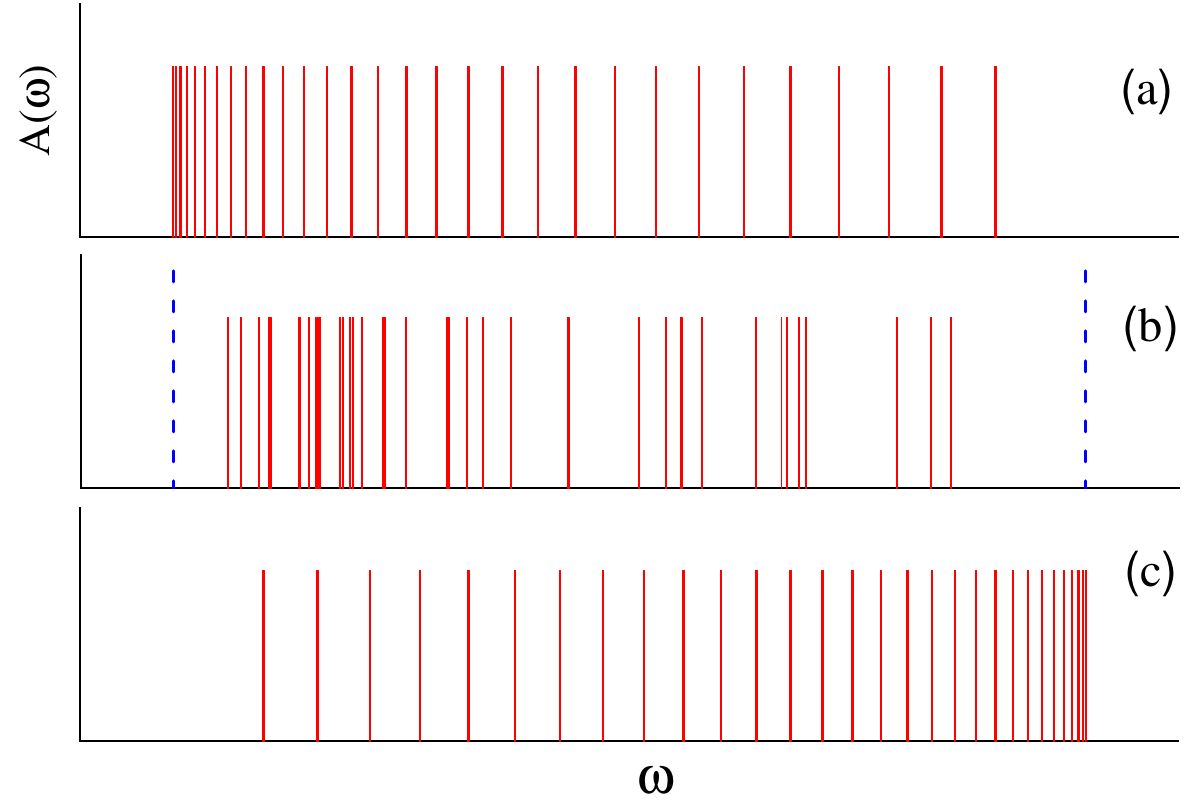}
\caption{Constraints on the sampled $\delta$-functions that are combined in various ways here in constrained SAC. In (a) and (c),
equal-weight $\delta$-functions are constrained to monotonically changing spacing, increasing from the left to right in (a) and vice versa in
(c); $\omega_i-\omega_{i-1}\leq \omega_{i+1}-\omega_i$ and $\omega_i-\omega_{i-1}\geq \omega_{i+1}-\omega_i$, respectively. In (b) the locations of
otherwise unconstrained $\delta$-functions are bounded from below by the edge in (a) or, for a two-edged spectrum, also from above by the edge
in (c). The relative weights of the different components of the spectrum can be optimized by $\chi^2$ minimization at fixed sampling temperature.
The locations of all the $\delta$-functions are sampled, including the edge(s).}
\label{Fig.edgecontpara}
\end{figure} 

It has been shown that a sharp edge followed by a continuum with power-law behavior (close to the edge) can be obtained in SAC when the distance
$d_i=\omega_{i}-\omega_{i-1}$ is constrained to be monotonically increasing with $i$ \cite{Shao23}, as illustrated in Fig.~\ref{Fig.edgecontpara}(a).
This constraint is associated with an entropic pressure forcing a divergent form (in the limit of a large number $N_\omega$ of $\delta$-functions)
$A(\omega) \sim (\omega -\langle \omega_1\rangle)^{-1/2}$, where $\langle \omega_1\rangle$, the edge location, is the mean of the energy of the lowest
$\delta$-function. The sampled edge location typically fluctuates very little once the process has equilibrated and $\omega_1$ has reached a small
optimal frequency window. Away from the immediate neighborhood of the edge, the spectrum can take any monotonically decaying form as dictated by the
QMC data. In the case of the prototypical Heisenberg chain, the dynamical spin structure factor (which exhibits a spinon continuum with a divergent spectral
weight at the lower edge) obtained this way compares very favorably with Bethe ansatz results \cite{Shao23}. To incorporate possible nonmonotonic
features above the edge, the constrained parametrization can be combined with another set of $\delta$-functions, which are unconstrained except
for the lowest frequency required to be above the edge $\omega_1$ of the distance-monotonic set. The relative weight of the two sets can be optimized by
minimizing the goodness of fit $\chi^2$ at a fixed sampling temperature; see Ref.~\cite{Shao23} for details.

Here we further extend this approach to also involve an upper edge, combining three different sets of mutually constrained $\delta$-functions, as
illustrated in Fig.~\ref{Fig.edgecontpara}. We will apply both the edge and double-edge constrained parametrizations next. While it is possible to
optimize the relative weights of all three sets (two different weight ratios), in test cases we find that the spectrum does not change very much when
the weights are of similar magnitude. This is because the unconstrained and constrained parts can adapt in many different ways to represent a near
optimal weight distribution. Therefore, to avoid excessively time consuming optimizations of a large number of spectra $A(k,\omega)$, we simply
set the relative weight of all three sets to $1/3$ for most cases.

Results with the double-edge constraint are shown in Fig.~\ref{Fig.t1J04logcombine}(b) and examples of fixed-$k$ profiles are shown in Fig.~\ref{Fig.sw_compare_all}(b).
Here the sharp bounds coincide very closely with the predicted edge singularities. The U-shaped broad band above the lower holon edge is now absent,
confirming our suspicion that this feature was spurious and a consequence of the insufficient resolution of the lower edge in the unconstrained results
in Fig.~\ref{Fig.t1J04logcombine}(a). The expected holon singularities inside the continuum are resolved as intermediate peaks at locations marked by
circles in Fig.~\ref{Fig.t1J04logcombine}(b) that are typically close to their predicted singular holon edges. Note that the same color scale is used for all
the heat maps in Fig.~\ref{Fig.t1J04logcombine}, which
implies that the bright broadened bands close to the expected edge are not present in the constrained results, where the edges are tall but very thin---the
fixed-$k$ profiles in Fig.~\ref{Fig.sw_compare_all} are clearer in this regard.

Given that the spin-charge separation ansatz is not exact and multi-spinon excitations in principle should also exist, there is likely some small
amount of spectral weight also above the upper singular edge. It is therefore also useful to consider only a constrained edge at the lower end of the
spectrum. These results are shown in Figs.~\ref{Fig.t1J04logcombine}(c) and \ref{Fig.sw_compare_all}(c). The lower edge does not move much relative to that
in the double-edge results, but other features are more smeared out. The upper edge is still sharper than in the completely unconstrained results, as is
the holon band away from the edges. The very small amount of weight above the location of the sharp upper bound in the double-edge results indicates that
there is only very little weight there also in the true spectrum---even smaller than what is seen in Fig.~\ref{Fig.t1J04logcombine}(c), considering the broadening
present in the SAC spectrum. The ED spectrum for $L=28$ \cite{Wrzosek24} also does not show substantial weight above the predicted upper edge.

\begin{figure}
\includegraphics[width=6.5cm]{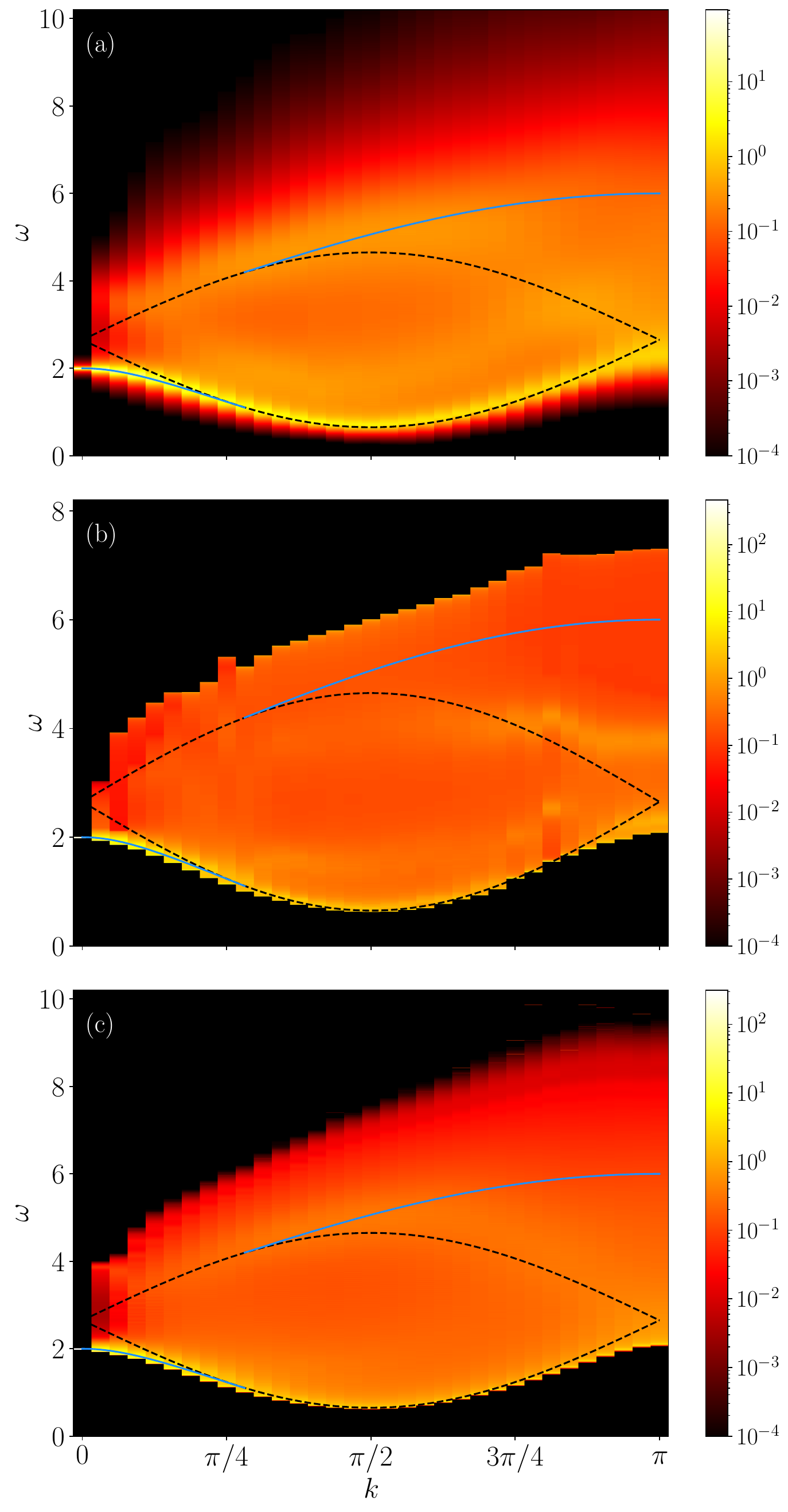}
\caption{Spectral function $A(k,\omega)$ of the $t$-$J$ model at the supersymmetric point, $t=1$, $J=2$. The panels correspond to three
different variants of SAC in the same way as in Fig.~\ref{Fig.t1J04logcombine} and the curves show the predicted locations of the singularities with
$J_s=1.35$ and $t_h=2$ in Eqs.~(\ref{disp}), chosen to match the holon dispersion close to $k=\pi$ and the putative spinon branch close to $k=0$.}
\label{Fig.t1J2combine}
\end{figure} 

\begin{figure}
\includegraphics[width=6.5cm]{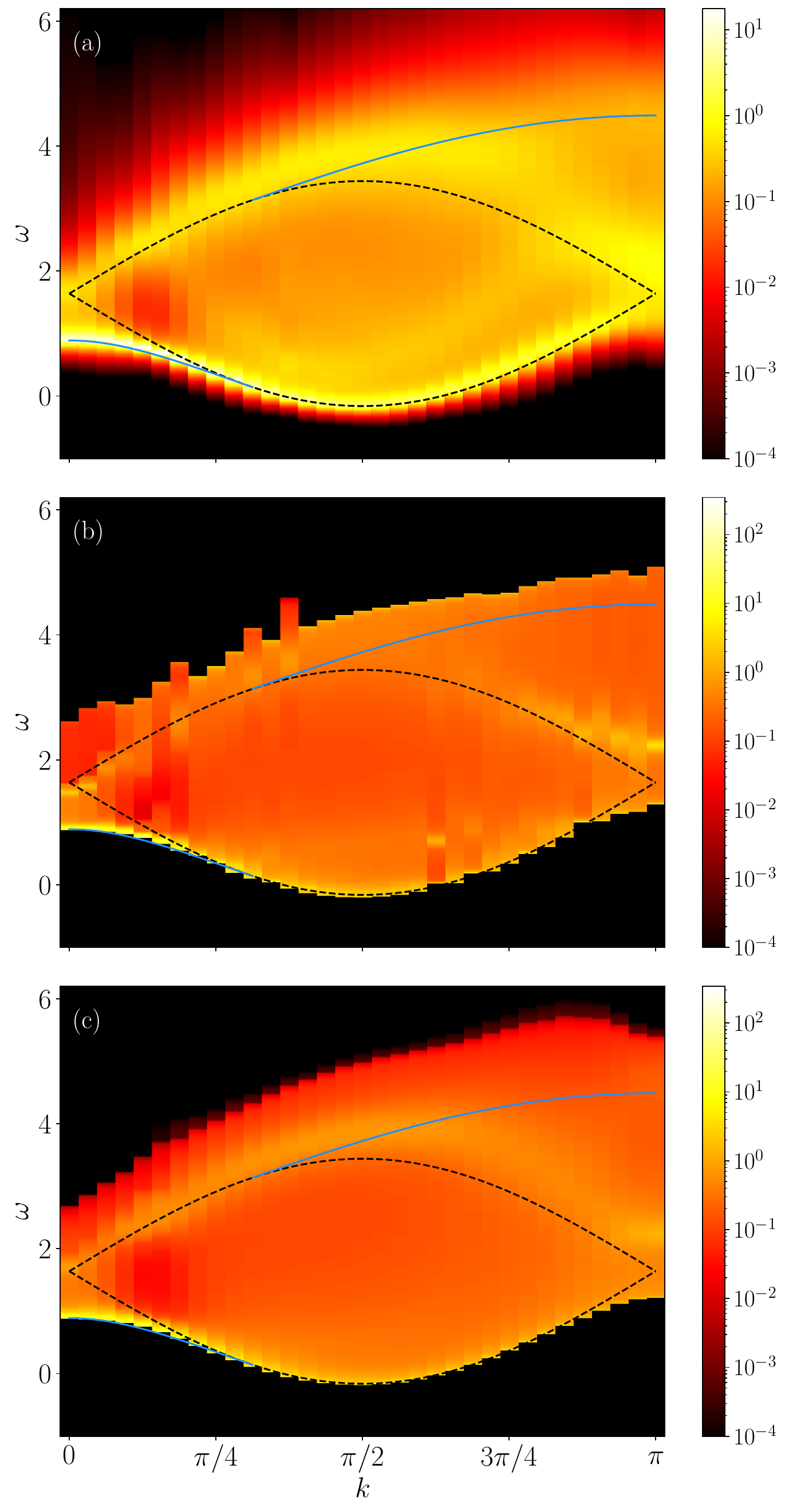}
\caption{Spectral function of the $t$-$J$-$Q$ model at the dimerization transition; $t=1$, $J=1$, $Q=0.1645$. The panels correspond to three
different variants of SAC in the same way as in Figs.~\ref{Fig.t1J04logcombine} and \ref{Fig.t1J2combine}. The curves show the predicted locations of the
singularities, with the holon hopping strength $t_h=1.8$ chosen to optimize the agreement with the lower edge close to $k=\pi/2$. The constant in the spinon 
dispersion Eq.~(\ref{disp_s}) is set at $J_s=1.05$ to match the lower edge (putative spinon branch) close to $k=0$.}
\label{Fig.t1J1Q01645logmorecombine}
\end{figure} 

We next consider the $t$–$J$ chain at the supersymmetric point, with $t=1$ and $J=2$ (thus, the off-diagonal spin interaction and the kinetic energy have the
same matrix elements), where the $k=0$ single-hole spectral function is exactly a $\delta$-function; $A(k=0,\omega)=\delta(\omega-2t)/2$ \cite{Sorella96,Sorella96_2,Sorella98}. As seen in Fig.~\ref{Fig.t1J2combine}(a), the unconstrained SAC result is fully consistent with this form, with only a small broadening arising from the statistical 
errors in imaginary time. In Figs.~\ref{Fig.t1J2combine}(b) and \ref{Fig.t1J2combine}(c), we have fitted $G(0,\tau)$ to a single $\delta$-function, and adding other
contributions does not improve the fit (a constrained parametrization that we will discuss in more detail in Sec.~\ref{sec:tjq_2}). Away from $k=0$, in the supersymmetric 
model there is no separate lower spinon branch within the spin-charge separation ansatz if the bare dispersion parameters $t_h=2t$ and $J_s=J$ are used, because then 
$k_0=0$. However, all three SAC parametrizations in Fig.~\ref{Fig.t1J2combine} consistently show a flattening out of the lower edge of the spectrum both for $k \to 0$
and $k \to \pi$. The entire lower edge, except close to $k=\pi$, can be well accounted for if we use different phenomenological spinon-holon parameters; here we only
adjust the spinon dispersion, with $J_s=1.35$, and keep the holon dispersion at its bare form. The deviations at $k=\pi$ cast some doubt on the predictions of the
spin-charge separation ansatz, as we will discuss further below.

Also in this case, with unconstrained SAC a likely spurious broad band is produced above the lower holon branch , which is seen more clearly in the $k=\pi/2$
profile in Fig.~\ref{Fig.sw_compare_all}(d). With the double-edge constraint applied in Fig.~\ref{Fig.t1J2combine}(b), the predicted upper edge
is not as well reproduced as in the previous $J/t=0.4$ case in Fig.~\ref{Fig.t1J04logcombine}(b). Moreover, the dim band that may be spurious still appears
within the predicted holon bounds. It is possible that the model with this rather large $J/t$ value has more substantial high-energy spectral weight that is
not captured by the spin-charge separation ansatz, which includes only one excited anti-spinon. The enforced upper edge would then be pushed up to partially
include this weight, which is possible since the statistical information in $G(k,\tau)$ is limited and SAC can produce good fits to the data even though the imposed behavior
at the upper edge is not strictly correct. Indeed, when only the lower edge is imposed, Fig.~\ref{Fig.t1J2combine}(c) shows a dim broad upper holon band close to 
the predicted location, and there is no spurious band close to the lower edge. Spectral weight extends to higher $\omega$ than in Fig.~\ref{Fig.t1J2combine}(b), 
well above the predicted upper bound. The difference between the predicted and observed bound looks smaller in the case of $J=0.4$, Fig.~\ref{Fig.t1J04logcombine}(c), 
for which the QMC data quality is very similar. Note, however, that the logarithmic heat map visually exaggerates regions of small but nonzero spectral weight, and 
the linear scales used in Fig.~\ref{Fig.sw_compare_all} are better suited for quantitative judgment.

Turning on the $Q$ interaction, in Fig.~\ref{Fig.t1J1Q01645logmorecombine}, as well as Figs.~\ref{Fig.sw_compare_all}(g)-\ref{Fig.sw_compare_all}(i), results
are shown for the three different SAC parametrizations in the case where $Q/J$ is set to the critical value of the dimerization transition of the host, where the 
marginal operator driving the transition vanishes. Here we have set the holon coefficient to $t_h=1.8$ so as to best reproduce the lower bound close to the minimum at
$k=\pi/2$. For the spinon dispersion, we set $J_s=1.05$ to reproduce the flattened edge close to $k=0$ in terms of a putative spinon band. Similar to the
supersymmetric point, there are clear deviations from the predicted holon lower bound close to $k=\pi$; here these deviations are even more pronounced.
Overall, the results do not differ much qualitatively from those for $Q=0$, which would also not be expected because the host spin system is still in the same type
of critical ground state, only without the marginal operator inducing logarithmic corrections.

Let us now return to the deviations from the predicted cosine form of the holon dispersion close to $k=\pi$ in both
Figs.~\ref{Fig.t1J2combine} and \ref{Fig.t1J1Q01645logmorecombine}. These deviations also cast some doubt on the matching of the edge close to $k=0$ by adjusting the phenomenological
dispersion parameters. The robust numerical observation is that the width of the holon spectrum does not shrink to a point as $k \to 0$ and $k \to \pi$
in the manner predicted by the mean-field spin-charge separation ansatz. This behavior is consistent with avoided level crossings of holon energies in the
extended Brillouin zone (BZ), since all states excited by the fermion destruction operator have the same quantum numbers (i.e., the total momentum, the parity, and the spin quantum numbers of the finite-size system). In principle such a gap may shrink to zero with increasing system size, but within our QMC resolution the lower edge with its flattening for $k\to 0$ and $k \to \pi$ is only very weakly dependent on
the system size for $L \ge 32$. This is not a full finite-size scaling analysis of the gap itself, but it supports the stability of the feature resolved
by QMC/SAC near the momenta where the ansatz predicts a crossing. We therefore regard the gap as evidence for interactions not included in the ansatz, while leaving its microscopic origin
as an open question for more refined analytical treatments. The ansatz-based fit to a lower spinon branch in Figs.~\ref{Fig.t1J2combine} and
\ref{Fig.t1J1Q01645logmorecombine} should therefore be viewed as suggestive rather than definitive; it is possible that the lower edge always corresponds
to the holon dispersion in these cases and that a separate spinon branch only forms for small $J/t$ in the $t$-$J$ model and also for small $Q/t$ in
the $t$-$J$-$Q$ model.

\subsection{Spontaneously dimerized phase}
\label{sec:tjq_2}

\begin{figure}
\includegraphics[width=6cm]{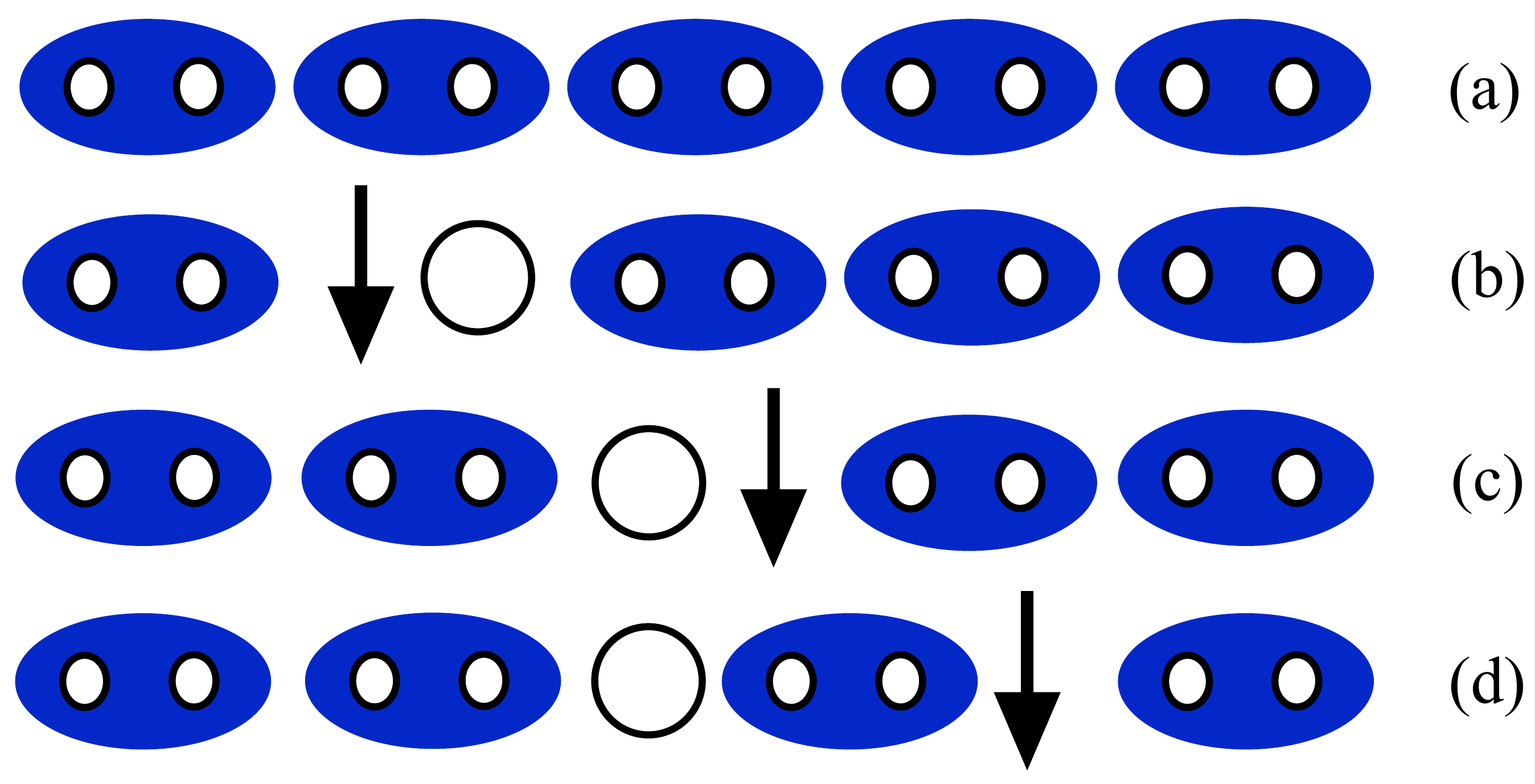}
\caption{Illustration of a single hole injected into the VBS phase (a), where blue ovals correspond to singlets. In (b) the resulting holon
is accompanied by a neighboring unpaired spin. The tightly bound spinon-holon object (spin-polaron) in can hop by quantum fluctuations and leave the VBS pattern
in the same phase A, exemplified in (c). The spinon and holon can also separate, as in (d), thereby creating a segment of singlets forming a translated 
VBS pattern B in between. Such states may either evolve to become fully spin-charge separated or contain extended spinon-holon bound states as discussed 
in the text.}
\label{Fig.t1JQmove}
\end{figure}

For $(Q/J) > (Q/J)_c \approx 0.1645$ the $J$-$Q$ chain is spontaneously dimerized \cite{Tang11,Yang20}, i.e., with a two-fold degenerate ground state,
$|A\rangle$ and $|B\rangle$, in the thermodynamic limit. Here spinons represent domain walls between $|A\rangle$ and $|B\rangle$ ordered chain segments
and they are also deconfined. It is not a priori clear whether residual interactions could cause binding of a holon and a spinon into a spin polaron,
however. We illustrate this question in Fig.~\ref{Fig.t1JQmove}, where the injection of a hole into a dimerized chain leads to the replacement of a
singlet by a hole and a spinon. The spinon-holon pair can propagate without separating, but once the partons do separate they may in principle
move further away and deconfine without additional energy cost. However, if there is a residual attractive interaction, one or several bound states
should form. The binding forces here include the loss of negative kinetic energy that allows the spinon and holon to resonate
on a single bond without moving. In an extreme case of a localized spin polaron surrounded by unbreakable singlets, the ground state energy of the
spinon-holon pair (in the even state with respect to spinon--holon permutation) is $-t$, whereas the corresponding internal excited state (odd with respect
to permutation) is $+t$. The cost of breaking up the pair is therefore $2t$, though this cost is compensated by the kinetic energy of the
deconfined partons and by the different effective spin-spin interactions in the neighborhood of the spinon and holon. Deconfined states will definitely
exist at sufficiently high energy, but it is possible that at least the lowest state could be a spin polaron for some momenta, which would be
manifested in $A(k,\omega)$ as an isolated $\delta$ function at energy $\omega_k$.

\begin{figure}
\includegraphics[width=7.5cm]{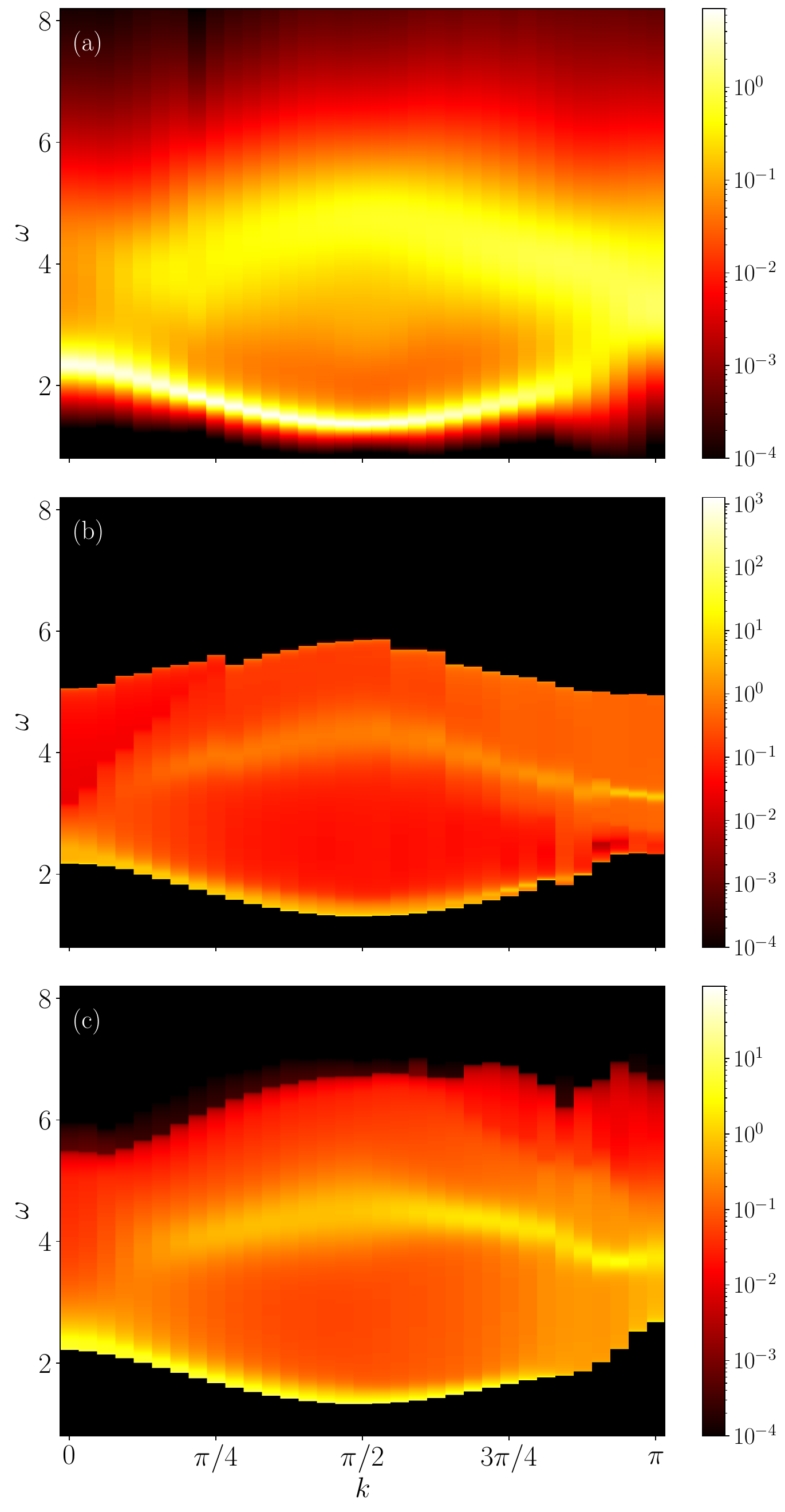}
\caption{Spectral function of the $t$-$J$-$Q$ model in the dimerized phase, with $t=1$, $J=0$, $Q=1$. The SAC parametrizations are the same
as in the previous 3-panel graphs.}
\label{Fig.t1J0Q1logcombine}
\end{figure} 

Before addressing the possibility of a spin polaron quasiparticle, in Fig.~\ref{Fig.t1J0Q1logcombine} we present results for $A(k,\omega)$ at $t=Q=1$, $J=0$, i.e.,
deep inside the VBS phase, using the same
three SAC parametrizations as in Sec.~\ref{sec:tjq_1}. While at first sight the results look very similar to the previous cases with the critical host state,
there are some notable differences. In particular, the band previously categorized as the upper holon band is much more prominent, and there is also 
more spectral weight above it. There is no sign of separate spinon and holon bands at small $k$ and the band width here is large compared to the previous
cases. For $k$ close to $\pi$, the shape of the spectral weight distribution is also different from previous cases, narrowing on approach to $k=\pi$ instead
of exhibiting the increase in the upper band energy that was due to the spinon branch in the previous cases. In the scenario of a spin polaron as the
lowest excitation, the bright second band in the middle of the continuum, which is particularly prominent in Fig.~\ref{Fig.t1J0Q1logcombine}(c), is
interpreted as the permutation-odd spin polaron, while the lower edge corresponds to the permutation even case.

\begin{figure}
\includegraphics[width=8cm]{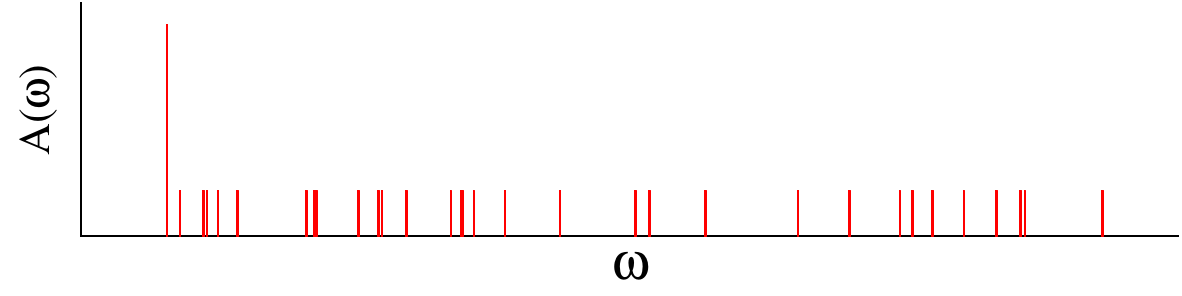}
\caption{Parametrization of a spectral function with a sharp quasiparticle edge modeled by a macroscopic $\delta$-function with relative weight $a_0$,
with its location $\omega_0$ being the lower bound for a large number of equal-amplitude microscopic $\delta$-functions representing a continuum with weight $1-a_0$.
The optimal value of $a_0$ is determined by a scan to find the best statistical match of the sampled average to the QMC data.}
\label{singlepeakpara}
\end{figure}

\begin{figure}
\includegraphics[width=7.5cm]{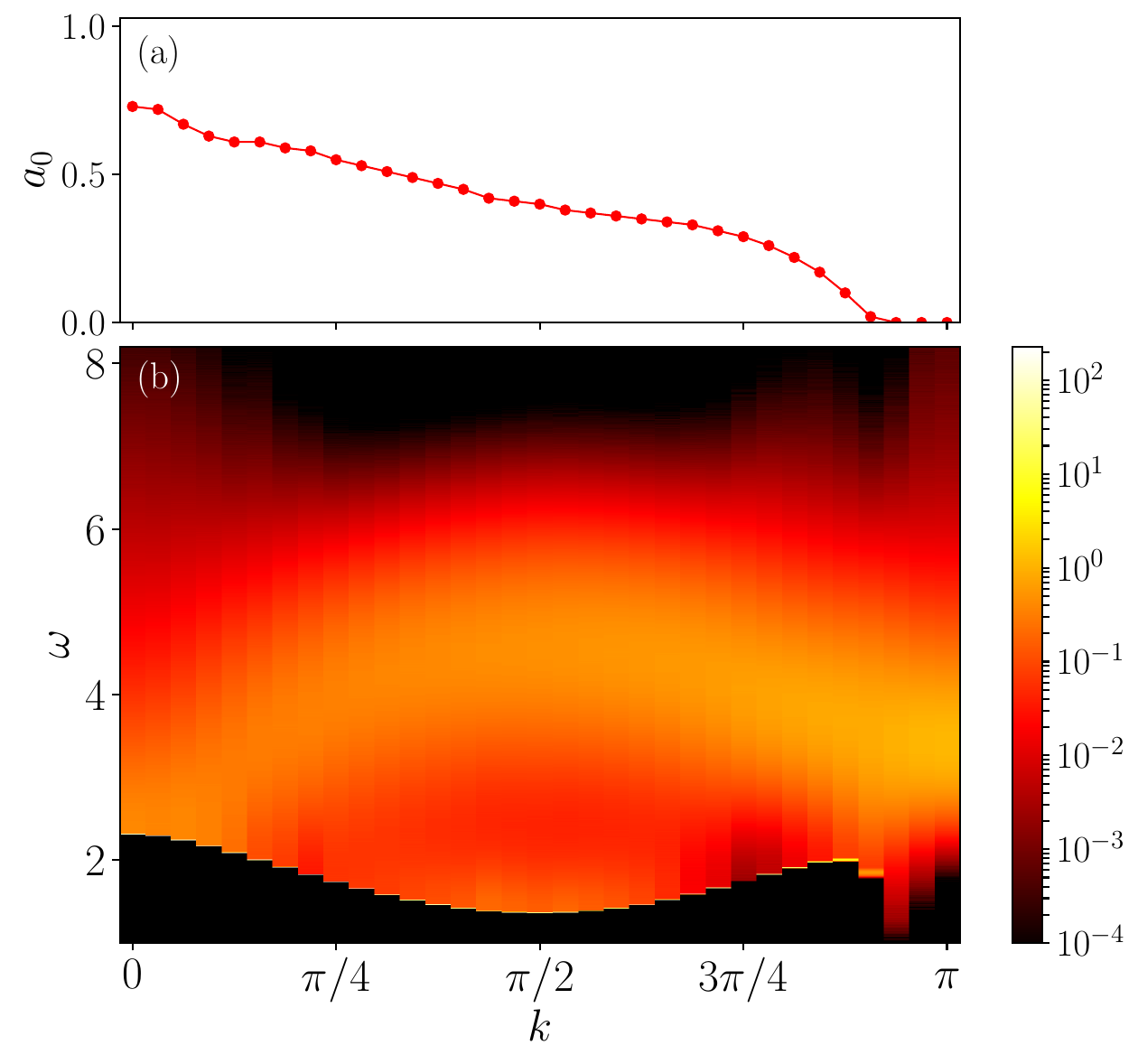}
\caption{Spectral function of the $t$-$J$-$Q$ model with $t=1$, $J=0$, and $Q=1$, using the parametrization with $\delta$-function edge
in Fig.~\ref{singlepeakpara}. The quasiparticle weight $a_0(k)$ is shown in (a). The heat map of $A(k,\omega)$ in (b) shows the contributions
above the edge, with relative weight $1-a_0(k)$. The small weight $a_0(k)$ close to $k=\pi$ makes the lower edge of the spectrum difficult to determine there, as reflected in the significant
scatter.}
\label{Fig.t1J0Q1SPa0All}
\end{figure}

To further investigate the spin polaron scenario, we adopt the SAC parametrization first introduced in Ref.~\cite{Shao17} and illustrated in
Fig.~\ref{singlepeakpara}, where a sharp quasiparticle at the lower edge of the spectrum is represented by a single ``macroscopic'' $\delta$-function
holding a fraction $a_0$ of the total spectral weight, followed by a continuum represented by a large number $N_\omega$ of equal-weight, $a_i=(1-a_0)/N_\omega$,
$\delta$-functions above $\omega_0$. The quasiparticle frequency $\omega_0$ is sampled at constant $a_0$ along with all $\omega_{i>0}$. Once the sampling
has equilibrated, $\omega_0$ is essentially fixed (unless $a_0$ is very small). 
The parameter quasiparticle weight is determined by optimizing the goodness-of-fit $\chi^2$ in a scan
over a range of $a_0$ values. Using this approach we obtain the momentum dependent spectral function and optimal quasiparticle weight $a_0(k)$
shown in Fig.~\ref{Fig.t1J0Q1SPa0All}. Notably, the quasiparticle weight is large close to $k=0$ and drops to zero in the neighborhood of $k=\pi$, in
the latter case causing large uncertainties in the actual location of the edge. The very small spectral weight at low energy for $k \approx \pi$ is
also seen with the power-law edge parametrizations in Figs.~\ref{Fig.t1J0Q1logcombine}(b) and \ref{Fig.t1J0Q1logcombine}(c). There is also no sign
of a spinon branch, either at the lower bound close to $k=0$ or at the upper bound of the spectrum close to $k=\pi$.

\begin{figure}
\includegraphics[width=8cm]{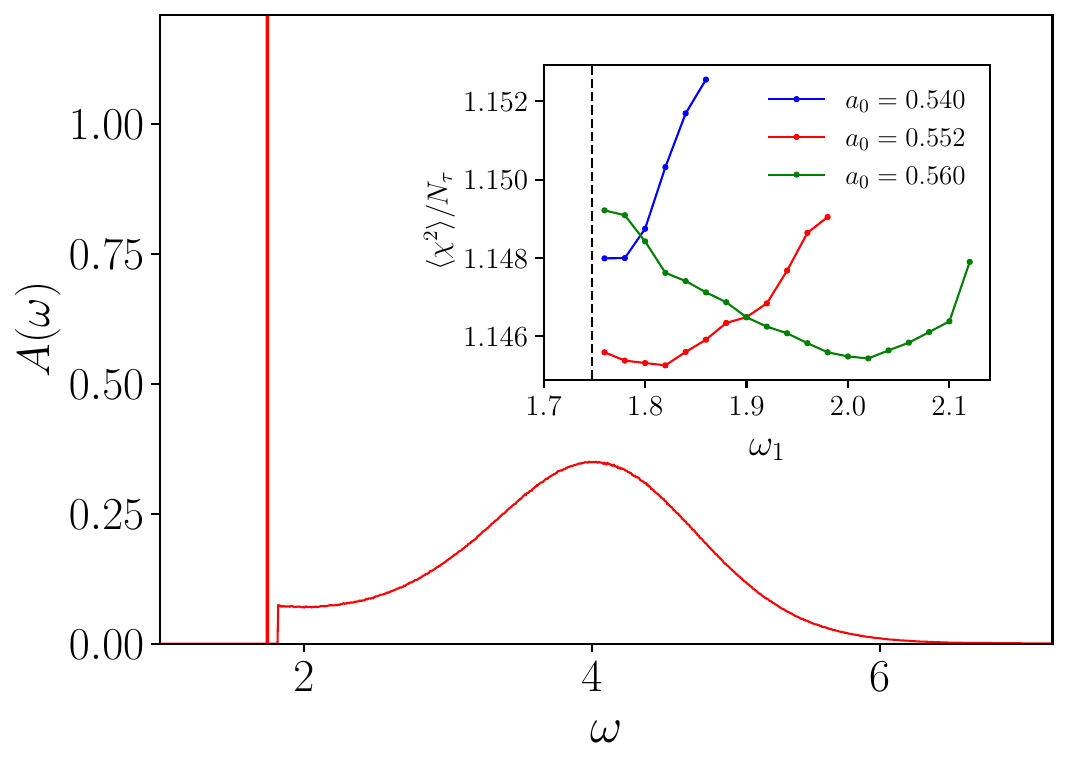}
\caption{Example of an optimized second gap with the same QMC data as in Fig.~\ref{Fig.t1J0Q1SPa0All} at $k=\pi/4$. The inset shows the goodness of fit versus the second gap $w_1$ for three
values of the weight $a_0$ of the edge $\delta$-function. The optimal parameters on the grid used are $a_0=0.552$ and $\omega_1=1.82$, and the vertical dashed line is the position of the lower bound $\omega_0$.}
\label{Fig.t1J0Q1SPa0w1}
\end{figure}

We do not detect any signs of a gap between the potential spin polaron and the continuum of excitations, though a very small gap cannot
be excluded. With the unconstrained sampling above the $\delta$-edge, it is not possible to resolve a very small gap because of entropic pressures
that tend to spread out the continuum while maintaining a good fit to the QMC data. To explore the possibility of a very small gap separating an
isolated sharp quasiparticle from the continuum, we impose a further constraint; a lower bound $\omega_1 > \omega_0$ for the contributions above the
macroscopic $\delta$-function at $\omega_0$. Here $\omega_0$ is again sampled but fluctuates very little, and $\omega_1$ is optimized along with the
weight $a_0$. To implement this optimization, we perform a scan over $\omega_1$ for several different values $a_0$,
searching for the optimal pair $(a_0, \omega_1)$ corresponding to the $\chi^2$ minimum. This optimization in a 2D parameter space is time consuming but helped by the fact that $a_0$ obtained without the additional
constraint is already very close to the further refined value. See Fig.~\ref{Fig.t1J0Q1SPa0w1} for a simple illustration where we demonstrate three $\omega_1$ scans and the corresponding optimal $A(\omega)$.
 
Using the second-gap approach, the optimized gaps obtained with the data in Fig.~\ref{Fig.t1J0Q1SPa0All} (in cases where the weight $a_0$ is
not close to $0$) are very small. The case of $k=\pi/4$ is shown in Fig.~\ref{Fig.t1J0Q1SPa0w1}. The small optimized gap and the absence of spectral
weight piling up close to the $\omega_1$ edge are consistent with a spin-polaron interpretation: a single isolated low-energy state followed by a
continuum of excitations that may include long-lived larger spinon-holon states whose broad peaks merge into a continuum. There should still also be
a high-energy continuum of fractionalized excitations.

These data do not by themselves constitute a proof of confinement in the spontaneously dimerized phase. The evidence for the spin-polaron scenario is
instead the combination of (i) the disappearance of clear separate spinon and holon bands at small $k$, (ii) the sizable optimized quasiparticle weight
away from $k\approx\pi$, and (iii) the qualitative agreement with bound-state physics in dimerized or frustrated chains. The presence of at least one
bound state has also previously been found in the frustrated $J_1$-$J_2$ Heisenberg chain at the special point $J_2/J_1=1/2$ where the ground state
is an exact dimer-singlet state \cite{Kuzian03,Tohyama98,Hayn00,Jurecka01}. This analogy strengthens, but does not by itself prove, the case for bound
states in the spontaneously dimerized $t$-$J$-$Q$ chain.

Within this interpretation, the lowest edge with a quasiparticle peak corresponds to the even (with respect to permutation of the partons) spin-polaron and
the broad band at higher energy in Fig.~\ref{Fig.t1J0Q1SPa0All} corresponds to an odd state dressed by multi-spinon excitations. The separation between the
even and odd cases should be of order $2t=2$, but is somewhat larger when maximized at $k=\pi/2$ in the results presented here. 

\section{Spin polarons in bond-alternating chains}
\label{sec:dimerized}

To further investigate spin polarons in 1D systems, we consider the $t$–$J$ chain with alternating strong ($t_1$, $J_1$) and weak ($t_2$, $J_2$) bond
parameters. In the limit where $t_1$, $J_1 \gg t_2$, $J_2$ and $t_i \ll J_i$, an inserted hole strongly binds to a neighboring spin on a strong bond to form
a localized spin polaron. Here the sketch in Fig.~\ref{Fig.t1JQmove}(a) still applies, with the shown pattern assumed to be the one minimizing the energy,
e.g., the singlets reside on the strong bonds. When inserting a hole, the accompanying unpaired spin will then also reside on the same bond, where the singlet
was broken. In this case there is no doubt that a bound state forms, and this spin polaron will move with an effective hopping $\propto t^2_2/J_1$ between the
strong bonds, corresponding to Figs.~\ref{Fig.t1JQmove}(b) and \ref{Fig.t1JQmove}(c). The spin polaron has two internal degrees of freedom; hence there will be
even and odd bands separated by a gap $\propto t_1$ (note, however, that there are corresponding completely conserved quantum numbers only at special momenta,
as we will discuss further below). The partons can still separate, as in Fig.~\ref{Fig.t1JQmove}(d), but in this case a string (in an analogy
with a pair of quarks bound by a gluon string) of high-energy singlets form between them, unlike the VBS phase where the couplings are uniform. Thus, we expect
a spectrum of equally spaced even-odd pairs of levels. At high energies the string should become increasingly unstable; thus the lifetime of the bound
state should become shorter and the corresponding spectral features broader.

\begin{figure}
 \includegraphics[width=7.5cm]{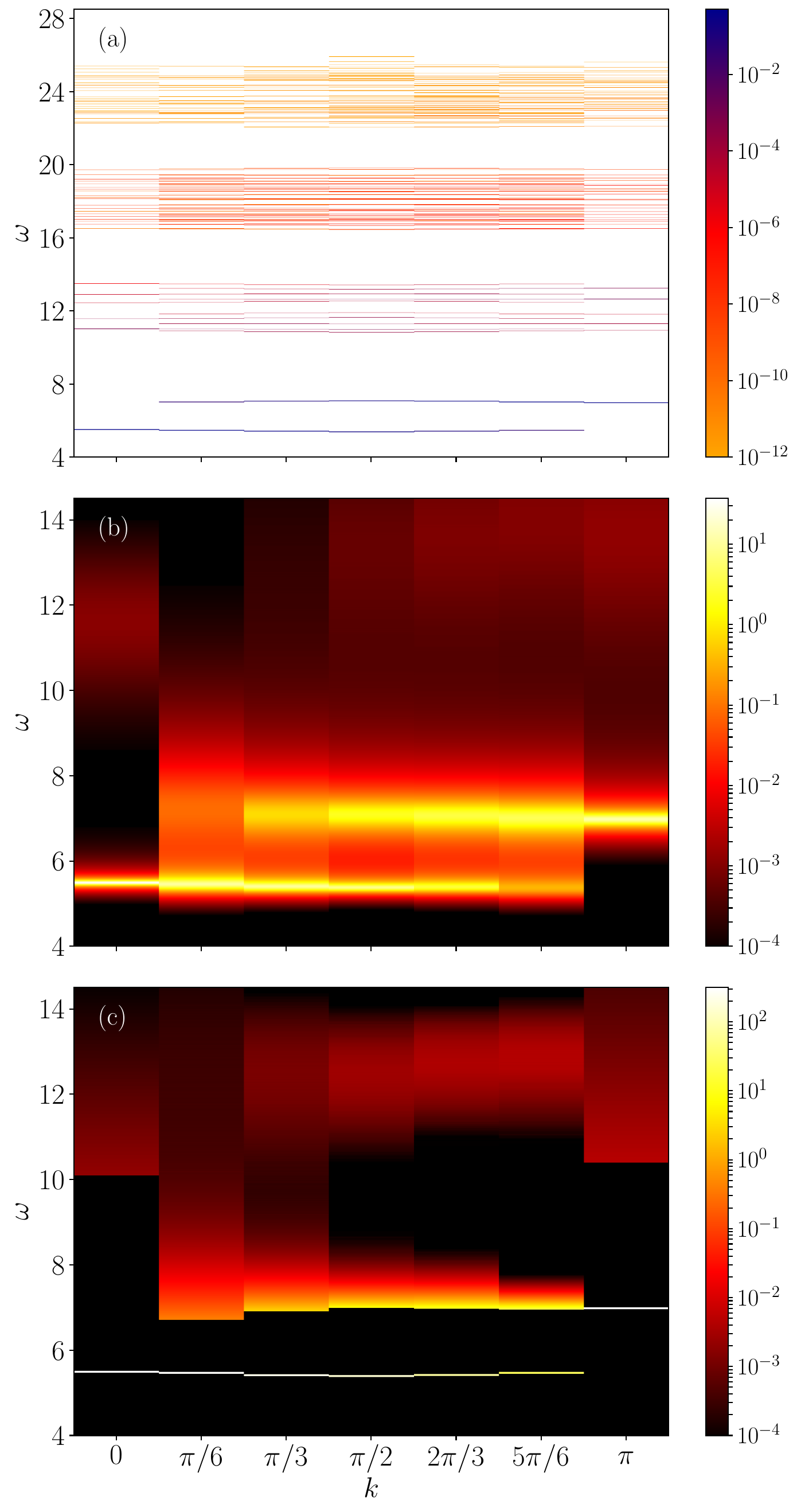}
\caption{Single-hole spectral function $A(k,\omega)$ of the bond-alternating $t_1$-$J_1$-$t_2$-$J_2$ chain of length $L=12$ with $t_1=0.8$, $J_1=6$, $t_2=0.1$,
$J_2=1$. (a) is based on full exact diagonalization, with the colors corresponding to the amplitudes of the $\delta$-functions in Eq.~(\ref{eq.singleholet0}).
The other panels show corresponding SAC results, with (b) from unconstrained sampling and (c) with a macroscopic $\delta$-function as the lower edge and an
optimized gap to the other contributions, i.e., sampling with the parametrization in Fig.~\ref{singlepeakpara} but with an additional optimized constraint
on the allowed space for the ``microscopic'' $\delta$ functions above the edge. Note that the color scale in (a) is inverted (darker colors for larger spectral
weights), to more clearly show the levels at the lowest energies.}
\label{Fig.Sw_EDL12_combine}
\end{figure}

To demonstrate this scenario explicitly, we first examine exact diagonalization (ED) results for $A(k,\omega)$ with parameters $t_1=0.8$, $J_1=6$,
$t_2=0.1$, $J_2=1$. Here we aim for a large number of low-energy levels and use full diagonalization for $L=12$, which is already sufficiently large for
displaying the low-energy level pattern used here as a benchmark, but not for a finite-size-converged spectrum. To present the results, we take the momentum from the unfolded BZ, i.e., that of the uniform chain, which makes it easier to analyze
different bands of excitations. Results for the energy levels versus momentum are shown in Fig.~\ref{Fig.Sw_EDL12_combine}(a), where the lines are colored
according to the squared matrix elements in the definition Eq.~(\ref{eq.singleholet0}) of the spectral function, using a logarithmic scale. The separation
between the lowest two levels is roughly $2t_1$ and the bands are only weakly dispersive on this scale, in agreement with the above reasoning.

A noteworthy feature is that the quasiparticle weight of the lowest band vanishes at $k=\pi$, while the second band shows no weight at $k=0$, even though the
corresponding states do exist in the spectrum. This behavior is consistent with the odd versus even internal permutation symmetry of the partons of the spin
polaron for the following arguments: $c_{k=0,\uparrow}$ creates an odd polaron when acting on the singlet product state illustrated in Fig.~\ref{Fig.t1JQmove}(a),
thus with no overlap with the even state. Conversely $c_{k=\pi,\uparrow}$ creates an even polaron with no overlap with the odd one. Since the sign structure of
the spin wave function (the Marshall sign rule) is maintained even if Fig.~\ref{Fig.t1JQmove}(a) is not the exact ground state, the arguments remain valid
also when $c_{0,\uparrow}$ and $c_{\pi,\uparrow}$ act on the true ground state of the spin model. Beyond these special cases, the even and odd branches are
not associated with a conservation law, but the internal parity of the spin polaron can still be long-lived and lead to the observed band formation.

The lowest level pair is separated from the next group of levels in Fig.~\ref{Fig.Sw_EDL12_combine}(a) (approximately in the frequency window
$\omega = 10 \sim 14$) by a gap consistent with the $J_1$ scale. This group corresponds to the second even and odd bound states, with emergent
broadening of these excitations corresponding to a finite lifetime of the higher bound states. Groups of levels at still higher energy are also broadened to
an increasing extent and the even and odd parts of the bands become difficult to distinguish. The separation between the bands is roughly constant in accord
with the linear potential of the string of mismatched singlets between the spinon and holon. While the lowest spin polaron can be
expected to remain a sharp excitation with a single $\delta$-function peak in $A(k,\omega)$, the sharp second band in Fig.~\ref{Fig.Sw_EDL12_combine}(a)
is likely a finite-size effect. For some model parameters and momenta, the second level indeed splits into two already for $L=12$. On increasing the excitation
energy, the string of dimerization mismatches will eventually break, which is facilitated by the creation of an additional spinon pair and subsequent
recombination of one spinon with the holon. The broadening, and loss of probability amplitude, of the higher bands in Fig.~\ref{Fig.Sw_EDL12_combine}(a)
reflect the short lifetime of the longer strings and the low probability of creating large bound states by hole injection.

Unconstrained SAC results for the same system are shown in Fig.~\ref{Fig.Sw_EDL12_combine}(b). The two lowest bands are resolved clearly, though
of course with some broadening. The spectral weight of the higher bands is very small and only a dim broad peak is resolved in the approximate
range of the second set of excitations around $\omega \approx 12$.
When using the parametrization with optimized macroscopic $\delta$-function edge, the dispersion relation of the
lowest band is well reproduced but there is still a small spectral weight between it and the second band. As shown in
Fig.~\ref{Fig.Sw_EDL12_combine}(c), we can completely separate the two lowest bands by imposing the second-gap constraint $\omega_1 > \omega_0$
for the contributions above the macroscopic $\delta$-function at $\omega_0$. Comparing with the ED results in 
Fig.~\ref{Fig.Sw_EDL12_combine}(a), $\omega_0(k)$ and $\omega_1(k)$ of the two lowest bands are very well reproduced in Fig.~\ref{Fig.Sw_EDL12_combine}(c);
within the same output bin width ($\Delta_\omega = $0.005) from SAC for $\omega_0(k)$ and to within at most $\sim$4\% for $\omega_1(k)$ for all $k$.

\begin{figure}
\includegraphics[width=\columnwidth]{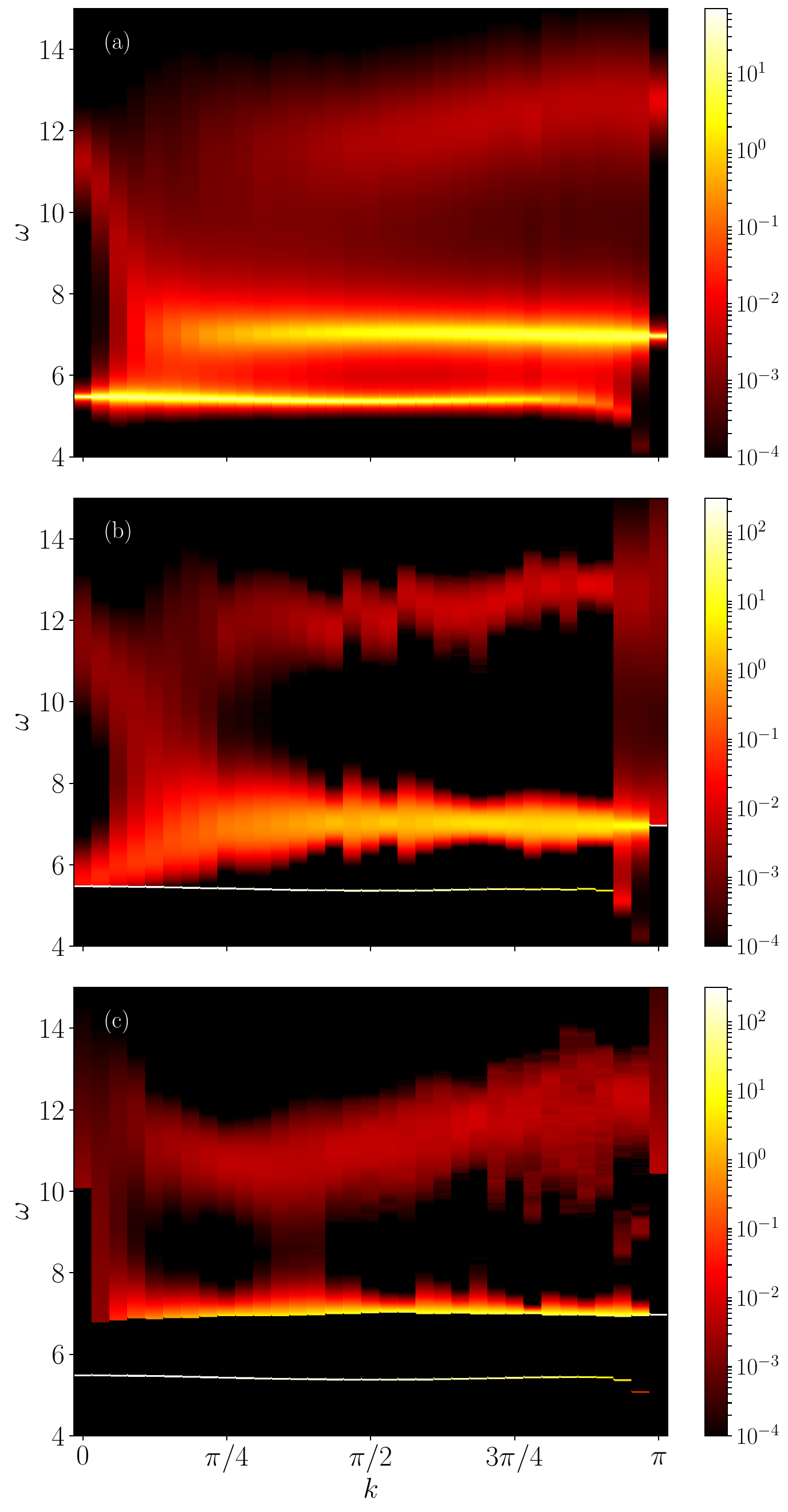}
\caption{Single-hole spectral function of the same model as in Fig.~\ref{Fig.Sw_EDL12_combine} but for system size $L=64$. Three different SAC parametrizations
are used as follows: (a) unconstrained with variable amplitudes, (b) single-$\delta$ quasiparticle edge (c) as in (b) but with an optimized lower bound for the
contributions above the first quasiparticle.}
\label{Fig.t108t201J16J21logcombine}
\end{figure}

\begin{figure}
\includegraphics[width=8cm]{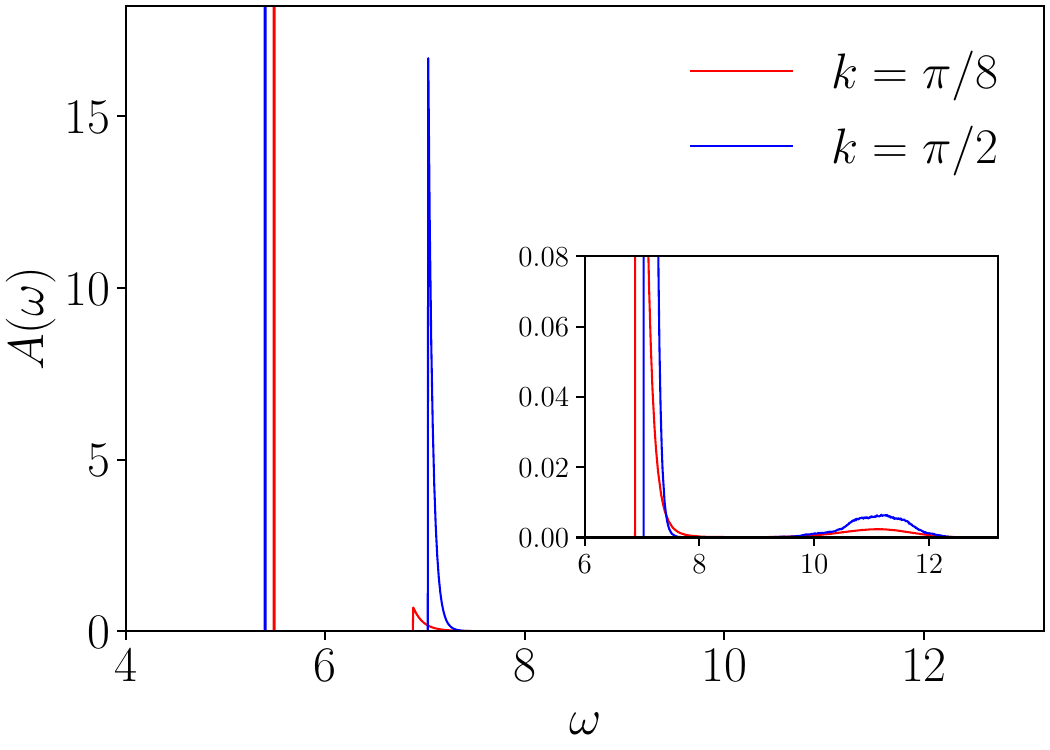}
\caption{Two representative vertical cuts of Fig.~\ref{Fig.t108t201J16J21logcombine}(c), at $k=\pi/8$ and $k=\pi/2$. The amplitude $a_0(k)$ of the lower
$\delta$-function was optimized along with the second gap $\omega_1(k)$ at which an asymmetric edge forms.
The inset is zoomed in on the $\omega_1$ edges and the small humps located at $\omega(k) = 10 \sim 12$.}
\label{Fig.swpi8pi2_dimer}
\end{figure}

\begin{figure}
\includegraphics[width=8cm]{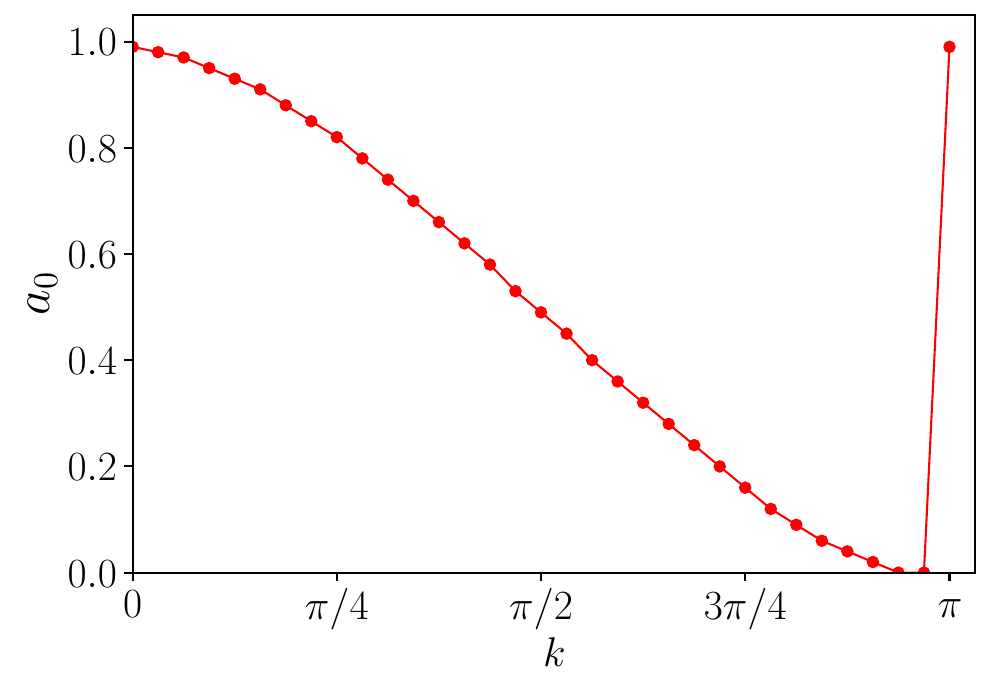}
\caption{Quasiparticle weight $a_0$ corresponding to Fig.~\ref{Fig.t108t201J16J21logcombine}(c). The jump at $k=\pi$ reflects vanishing
$a_0$ of the first band at this momentum and the method then identifies the quasiparticle of the second band.}
\label{Fig.t108t201J16J21a0All}
\end{figure}

Extending to larger system sizes with QMC and SAC, we observe overall similar features, as shown in Fig.~\ref{Fig.t108t201J16J21logcombine}
for the same parameter set as above but now for chain length $L=64$. Unconstrained SAC again reveals two narrow bands close to those for $L=12$ but with
more clearly visible gradual fading of the spectral weight for the lower band when $k \to \pi$ and for the second band when $k \to 0$. The third band is
better defined than in the $L=12$ unconstrained SAC results in Fig.~\ref{Fig.t108t201J16J21logcombine}(a) and is also more dispersive.
In Fig.~\ref{Fig.t108t201J16J21logcombine}(b), where the $\delta$-edge parametrization was used, the second band is better separated from the now
sharp lower band, except in the neighborhood of $k=0$ and $k=\pi$---in the latter case the very small spectral weight makes it difficult to determine the
optimal $a_0$ value.

Thus, in this bond-alternating case the ED spectrum in Fig.~\ref{Fig.Sw_EDL12_combine}(a) should be viewed as a small-system reference for the parity
structure and for benchmarking the SAC parametrizations. The $L=64$ QMC/SAC spectra provide the main
large-system comparison, but the regions near $k=0$ and $k=\pi$ remain the most delicate because one of the two lowest branches has vanishing or very
small spectral weight. We therefore focus on features that persist from the ED benchmark to
the larger QMC/SAC system, especially the two lowest spin-polaron bands and their transfer of spectral weight.

Further improving the results by optimizing the second gap $\omega_1(k)$, Fig.~\ref{Fig.t108t201J16J21logcombine}(c) also shows a rather sharp second band.
The distance between the lowest bands is again close to $2t_1$ as expected for the odd and even spin polarons at this large modulation of the couplings. An
interesting feature is that the third band (which likely involves merged even and odd sub-bands) now has a minimum between $k=\pi/4$ and $\pi/2$, unlike
the almost dispersionless band found above with ED for $L=12$. There is
very little spectral weight in a continuum at higher energies, reflecting the higher bound states that cannot be resolved. Fig.~\ref{Fig.swpi8pi2_dimer} shows
the spectra at two values of $k$ versus $\omega$ to further demonstrate the capability of the method with a second gap to resolve the two lowest bands and
also likely give a good approximation of the third band. The quasiparticle weight $a_0$ is shown versus $k$ in Fig.~\ref{Fig.t108t201J16J21a0All} based on
the results in Fig.~\ref{Fig.t108t201J16J21logcombine}(c). The $a_0$ values from Figs.~\ref{Fig.t108t201J16J21logcombine}(a) and \ref{Fig.t108t201J16J21logcombine}(b)
are very similar to those graphed. Note that the jump at $k=\pi$ reflects the absence of weight in the lower band at this
momentum, and the method then instead finds the $\delta$-edge in the second band. Since the spectral weight beyond the two lowest bands is very small,
the relative weight $a_1(k)$ of the second band is very close to $1-a_0(k)$; thus spectral weight is gradually transferred from the odd to the even
spin polaron as $k$ changes from $0$ to $\pi$.

\section{Summary and discussion}\label{sec:summary}

We have systematically investigated the fermionic single-hole spectral function $A(k,\omega)$ of $S=1/2$ quantum spin chains with different types of
ground states. Our main aim here has been to test the capabilities of constrained SAC to resolve sharp edge features that typically cannot be reproduced
with conventional analytic continuation methods---here exemplified by unconstrained SAC, although the maximum-entropy method \cite{Jarrell96} also has very similar
limitations. One may argue that the constraints introduce bias and that the approach either just confirms previously known facts or produces misleading results.
The reality, however, is that sharp features cannot be resolved unless some corresponding information is provided, and the imposed constraints improve the
ability of SAC (or other methods where similar constraints could be implemented) to deliver reliable dispersion relations for the lowest excitations,
reduce the presence of spurious features at higher energy, and better resolve actual features above the edge. Thus, when the existence of a sharp edge is
known, its incorporation as a constraint can be of great value. Note that the default models used in the maximum-entropy method could also in principle
be optimized in a way analogous to the constraints in SAC (see Ref.~\cite{Shao23} for further discussion) but SAC often represents a more natural
framework in this regard since the location of, e.g., the lower edge does not have to be known but is found by the method itself.

Here we have used both a lower edge with a sharp $\delta$-function and one corresponding to a divergent continuum, in
one case also using such an edge at the upper bound of the spectrum. We also demonstrated that a $\delta$-function edge followed by an optimized second
gap is useful for resolving additional low-energy bands. In cases where the exact nature of the edge is not known, it is useful to compare different constraints.
Here we have only provided visual comparisons, but in principle statistical cross-validation methods can also be used to single out the most likely
spectrum out of a group \cite{Schumm25}.

As for the results, in the case of the conventional $t$-$J$ chain and $t$–$J$–$Q$ chain with $Q/J$ inside the same critical phase of the host system,
the incorporation of power-law divergent edges produced spectral functions with some features closely matching the spin-charge separation ansatz. However, 
in some cases we also pointed out that the holon dispersion relation exhibits gaps between the upper and lower bounds at the point where the two should 
merge (cross each other in momentum space) according to the ansatz. This manifestation of the approximate nature of the ansatz motivates further study of spin-charge
separation with more sophisticated analytical methods. With large $Q/J$, inside the spontaneously dimerized phase, our results point to an isolated spin polaron
bound state as the lowest excitation. This state should be even with respect to parton permutation and we also detected a higher band consistent with
odd-parity polarons with a finite lifetime. In the statically dimerized chain, we were likewise able to resolve multiple bands of spin polarons.

The exact-diagonalization results used here should be understood as benchmarks for small systems and for the QMC estimator rather than as a complete
finite-size analysis. The large-system conclusions rely on QMC/SAC spectra for $L=64$ chains. For the uniform-chain holon-gap issue, tests indicate
that the lower edge with its flattening for $k\to0$ and $k\to\pi$ is only weakly dependent on the system size for $L \ge 32$. We have therefore been
cautious in discussing the most delicate momenta, especially near $k=0$ and $k=\pi$, where small spectral weights and possible avoided crossings make
the interpretation less direct.

Overall, our results demonstrate the effectiveness of constrained SAC approaches in reproducing spectral functions with sharp
features that are beyond the resolution of other numerical methods. In combination with large-scale quantum Monte Carlo simulations, these SAC methods
highlight the potential for much broader applications to a wide range of quantum many-body systems. Results for the quantum-critical 2D $t$-$J$-$Q$
model are reported in Ref.~\cite{Yang25new2}.

\begin{acknowledgments}
We would like to thank Karlo Penc and Qimiao Si for discussions. This research was supported by the Simons Foundation under Grant No.~511064.
The numerical calculations were carried out on the Shared Computing Cluster managed by Boston University’s Research Computing Services.
\end{acknowledgments}

\section*{Data Availability}
The data that support the findings of this article are not publicly available upon publication because it is not technically feasible and/or the cost of
preparing, depositing, and hosting the data would be prohibitive within the terms of this research project. The data are available from the authors upon
reasonable request.

\bibliographystyle{apsrev4-2-with-titles}
\bibliography{1dtjq.bib}

\end{document}